\documentclass[useAMS]{mn2e}
\usepackage{graphicx} 
\usepackage{subfig}
\usepackage{amssymb}
\usepackage{natbib}
\usepackage{amsmath}
\usepackage{multirow} 
\usepackage{booktabs,threeparttable}
\usepackage{color}
\usepackage{graphicx,url,times}
\usepackage{array}
\DeclareGraphicsExtensions{.ps,.pdf,.png}
\makeatletter
\def\fps@figure{htbp}
\makeatother

\newcolumntype{P}[1]{>{\raggedleft\arraybackslash}p{#1}}

\begin{document}

\title[Galaxy Cluster Mass Reconstruction]{Galaxy Cluster Mass Reconstruction Project: \\
I. Methods and first results on galaxy-based techniques}

\author[Old et al.]{L. Old$^{1}$\thanks{E-mail:
ppxlo@nottingham.ac.uk}, R. A. Skibba$^{2}$, F. R. Pearce$^{1}$,
D. Croton$^{3}$, S. I. Muldrew$^{1}$,\newauthor
J. C. Mu\~{n}oz-Cuartas$^{4}$, D. Gifford$^{5}$, M. E. Gray$^{1}$, A. von der Linden$^{6,7,8}$,
G. A. Mamon$^{9}$, \newauthor M.R. Merrifield$^{1}$,
V. M\"uller$^{10}$, R. J. Pearson$^{11}$, T. J. Ponman$^{11}$, A. Saro$^{12}$,
T. Sepp$^{13}$,\newauthor C. Sif\'on$^{14}$, E. Tempel$^{13,15}$, E. Tundo$^{1}$,
Y. O. Wang$^{1,16,17}$ and R. Wojtak$^{6}$ \\
$^{1}$School of Physics and Astronomy, University of Nottingham, Nottingham, NG7 2RD, UK\\ 
$^{2}$Center
for Astrophysics and Space Sciences, Department of Physics, University
of California, 9500 Gilman Dr, San Diego, CA 92093, USA\\ 
$^{3}$Centre
for Astrophysics \& Supercomputing, Swinburne University of
Technology, PO Box 218, Hawthorn, VIC 3122, Australia\\ 
$^{4}$Group for Computational Physics and Astrophysics, Instituto de
Fisica, Universidad de Antioquia, Medellin, Colombia\\
$^{5}$Department of Astronomy, University of Michigan, 500 Church
St. Ann Arbor, MI, USA\\ 
$^{6}$Dark Cosmology Centre, Niels Bohr Institute, University of Copenhagen, Juliane Maries Vej 30, DK-2100 Copenhagen Ã, Denmark\\ 
$^{7}$Kavli Institute
for Particle Astrophysics and Cosmology, Stanford University, 452
Lomita Mall, Stanford, CA 94305-4085, USA\\ 
$^{8}$Department of Physics, Stanford University, 382 Via Pueblo Mall, Stanford, CA 94305-4060, USA\\
$^{9}$Institut
d'Astrophysique de Paris (UMR 7095 CNRS $\&$ UPMC), 98 bis Bd Arago, F-75014 Paris, France\\
$^{10}$Leibniz-Institut f\"ur Astrophysik Potsdam, An der Sternwarte 16, D-14482 Potsdam, Germany\\ 
$^{11}$School
of Physics and Astronomy, University of Birmingham, Birmingham, B15
2TT, UK\\ 
$^{12}$Department of Physics,
Ludwig-Maximilians-UniversitÂ¨at, Scheinerstr. 1, D-81679
MÂ¨unchen, Germany\\ 
$^{13}$Tartu Observatory, Observatooriumi 1, 61602 T\~oravere,
Estonia\\ 
$^{14}$Leiden Observatory, Leiden University,
P.O. Box 9513, NL-2300 RA Leiden, The Netherlands\\
$^{15}$National Institute of Chemical Physics and Biophysics, R\"avala pst 10, Tallinn 10143, Estonia\\
$^{16}$Key Laboratory for Research in Galaxies and Cosmology, Shanghai Astronomical Observatory, Shanghai 200030, China\\
$^{17}$Graduate School, University of the Chinese Academy of Sciences, 19A, Yuquan Road, Beijing, China\\}

\date{Accepted ??. Received ??; in original form ??}
\pagerange{\pageref{firstpage}--\pageref{lastpage}} \pubyear{2013}
\maketitle

\label{firstpage}

\begin{abstract}
 This paper is the first in a series in which we perform an
 extensive comparison of various galaxy-based cluster mass estimation
 techniques that utilise the positions, velocities and colours of
 galaxies. Our primary aim is to test the performance of these cluster mass estimation techniques on a diverse set of models that will increase in complexity. We begin by providing participating methods with data from a simple model that delivers idealised clusters, enabling us to quantify the underlying scatter intrinsic to these mass estimation techniques. The mock catalogue is based on a Halo Occupation Distribution (HOD) model that assumes spherical Navarro, Frenk and White (NFW) haloes truncated at $R_{\rm 200}$, with no substructure nor colour segregation, and with isotropic, isothermal Maxwellian velocities. We find that, above $\rm 10^{14} M_{\odot}$, recovered cluster masses are correlated with the true underlying cluster
 mass with an intrinsic scatter of typically a factor of two. Below
 $\rm 10^{14} M_{\odot}$, the scatter rises as the number of member
 galaxies drops and rapidly approaches an order of magnitude. We find
 that richness-based methods deliver the lowest scatter, but it is not clear whether such
 accuracy may simply be the result of using an over-simplistic model to populate the galaxies in their haloes. Even when given the true cluster membership, large scatter is
 observed for the majority non-richness-based approaches, suggesting that mass
 reconstruction with a low number of dynamical tracers is inherently
 problematic.
\end{abstract}


\clearpage
\section{Introduction}
Deducing the masses of the largest gravitationally bound structures in
the Universe, galaxy clusters, remains a complex problem that is at
the focus of current and future cosmological studies. The
characteristics of the galaxy cluster population provide crucial
information for studies of large scale-structure (e.g.,
\citealt{1988ARA&A..26..631B}; \citealt{2001AJ....122.2222E};
\citealt{2005MNRAS.357..608Y}; \citealt{2008ApJ...676..206P};
\citealt{2013MNRAS.430..134W}), constraining cosmological model
parameters (see \citealt{2011ARA&A..49..409A} for a review) and
galaxy evolution studies (e.g., \citealt{2003MNRAS.346..601G};
\citealt{2005ApJ...623..721P}; \citealt{2008MNRAS.391..585M}). Despite
the wealth of information clusters can provide, deriving strong
constraints from cluster surveys is a non-trivial problem due to the
complexity of estimating accurate cluster masses. The use of cluster surveys as a dark energy probe provides greater statistical power than other techniques (Dark Energy Task Force; \citealt{2006astro.ph..9591A}). However, enabling this statistical precision requires significant advances in treating the systematic uncertainties between survey observables and cluster masses.\\
\indent Clusters can be detected across several different wavelength regimes
using various techniques. They are identified in optical and infrared
light as over-densities in the number counts of galaxies (e.g.,
\citealt{1958ApJS....3..211A}, \citealt{1968cgcg.book.....Z}), while
colour information improves the contrast by selecting the red galaxies
that dominate in these systems (e.g., \citealt{2005yCat..21570001G},
\citealt{2007ApJ...660..221K}, \citealt{2011ApJ...736...21S},
\citealt{2012MNRAS.420.1167A}). At X-ray wavelengths, the hot
intra-cluster medium produces bright extended sources (e.g.,
\citealt{1972ApJ...178..309F}, \citealt{2000ApJS..129..435B},
\citealt{2002ARA&A..40..539R}, \citealt{2009ApJ...692.1033V}), while
at millimeter wavelengths, inverse Compton scattering of photons from this gas
results in characteristic distortions in the cosmic microwave
background (e.g., \citealt{1972CoASP...4..173S},
\citealt{2002ARA&A..40..643C}, \citealt{2013arXiv1303.5089P},
\citealt{2010ApJ...722.1180V}, \citealt{2013JCAP...07..008H}).
Finally, distortions of images of faint background galaxies through
weak gravitational lensing offers perhaps the most direct measure of
the huge masses of these systems (e.g., \citealt{2012arXiv1208.0605A}).

\indent Despite these diverse methods of detecting clusters, no cluster observable \emph{directly} delivers a mass. The cluster mass function is one key method to constrain the dark energy parameter. Ongoing and future dark energy missions plan to consider cluster counts in their analyses. Hence, it is crucial to be able to measure cluster masses as accurately as possible. Follow-up spectroscopy is of great importance to all group/cluster surveys, providing the kinematics of cluster galaxies, which is one of a few mass proxies that is directly related to cluster mass (by providing a direct measure of the dark matter potential well). This series of papers examines various observable - mass relations by testing an extensive range of galaxy-based cluster mass estimation techniques with the aim of calibrating follow-up mass proxies.

\indent Galaxy-based mass estimation techniques commonly follow three general steps: first
identify the cluster overdensity, second deduce cluster membership,
and finally, using this membership, estimate a cluster mass. Common
optical cluster finding methods include using the
\citet{1982ApJ...257..423H} Friends-Of-Friends (FOF) group-finding
algorithm (e.g., \citealt{2006ApJS..167....1B};
\citealt{2008AJ....135..809L}; \citealt{2012A&A...540A.106T};
\citealt{2013arXiv1305.1891J}) and methods based upon Voronoi
tessellation (e.g., \citealt{2002ApJ...580..122M};
\citealt{2004AJ....128.1017L}; \citealt{2009MNRAS.395.1845V};
\citealt{2011ApJ...727...45S}). Also widely used are red-sequence
filtering techniques (e.g., \citealt{2000AJ....120.2148G};
\citealt{2012MNRAS.420.1861M}; \citealt{2013arXiv1303.3562R}) and
methods that rely on the bright central galaxy (BCG) to identify the
presence of a cluster (e.g., \citealt{2005MNRAS.356.1293Y};
\citealt{2007ApJ...660..221K}). Cluster catalogues are also
constructed using the positions and magnitudes of galaxies to search
for over-densities via the matched filter algorithm (e.g.,
\citealt{1996AJ....111..615P}; \citealt{1999A&A...345..681O};
\citealt{1999ApJ...517...78K}; \citealt{2009ApJ...698.1221M}).\\
\indent Once the over-densities are identified, many methods select an initial
cluster membership using the groups obtained via the FOF algorithm (e.g., \citealt{2012MNRAS.423.1583M}; Pearson et al. in preparation; \citealt{2012A&A...540A.106T}),
whilst others select galaxies within a specified volume in phase space
(e.g., \citealt{2007MNRAS.379..867V}; \citealt{2009MNRAS.399..812W};
\citealt{2013MNRAS.429.3079M}; \citealt{2013ApJ...768L..32G};
\citealt{2013ApJ...772...25S}; Pearson et al. in preparation) or within a
certain region of colour--magnitude space where cluster galaxies are
known to reside (e.g., \citealt{2013ApJ...772...47S}). Once the initial set of member galaxies is chosen, it is common to
iteratively refine membership using either the estimated velocity
dispersion, radius and colour information, or even a combination of
these properties. Deducing which galaxies are members of a cluster is
non-trivial, and unfortunately the inclusion of even quite small
fractions of interloper galaxies that are not gravitationally bound to
the cluster can lead to a strong bias in velocity dispersion-based mass estimates (e.g.,
\citealt{1983MNRAS.204...33L}; \citealt{1997NewA....2..119B};
\citealt{1997ApJ...485...39C}; \citealt{2006A&A...456...23B};
\citealt{2007A&A...466..437W}). For this reason, methods often employ
careful interloper removal techniques, for example, by modelling
interloper contamination when performing density fitting, by using the
Gapper technique or via iterative clipping as described above.\\
\indent Many methods then follow the classical approach of applying the virial
theorem to the projected phase space distribution of member galaxies
(e.g., \citealt{1937ApJ....86..217Z}; \citealt{1977ApJ...214..347Y};
\citealt{Evrard:2008vo}), assuming that the system is in 
equilibrium. Other methods utilise the distribution of galaxies in
projected phase space, assuming a Navarro, Frenk and White (NFW)
density profile (\citealt{1996ApJ...462..563N};
\citealt{1997ApJ...490..493N}) to obtain an estimate of cluster
mass. The number of galaxies associated with a cluster above a given
magnitude limit (the richness) is also used as a proxy for
mass (e.g., \citealt{2003ApJ...585..215Y}). 
In addition, the more-recently developed caustic method
identifies the projected escape velocity profile of a cluster in
radius-velocity phase space, delivering a measure of cluster mass
(e.g., \citealt{1997ApJ...481..633D}; \citealt{1999MNRAS.309..610D};
\citealt{2013ApJ...768L..32G}).\\
\indent The aim of this paper is to perform a comprehensive comparison of 23
different methods that employ variations of the techniques described
above by deducing both the mass and membership of galaxy clusters from a
mock galaxy catalogue. In order to simplify the problem, the clusters
are populated with galaxies in a somewhat idealised manner, with
cluster locations that are specified {\it a priori}; in this way, the basic
workings of the various algorithms can be tested under optimal
conditions, without the potential for confusion from more complex
geometries or misidentified clusters.\\
\indent The paper is organised as follows. 
We describe the mock galaxy catalogue in Section~2, and the
mass reconstruction methods applied to this catalogue are described in Section~3. 
In Sections~4 and 5, we present our results on cluster mass and membership comparisons. 
We end with a discussion of our results and conclusions in Section 6. 
Throughout the paper we adopt a Lambda cold
dark matter ($\Lambda$CDM) cosmology with $\Omega_{\rm 0}=0.25$,
$\Omega_{\rm \Lambda}=0.75$, and a Hubble constant of $H_{\rm 0} =
73\,\rm{km\,s^{-1}}\,\rm{Mpc^{-1}}$, although none of the conclusions depend
strongly on these parameters.

\section{Data}
\label{sec:Sim Data}
This paper forms the initial part (Phase I) of a large comparison
programme aimed at studying how well halo masses can be recovered
using a wide variety of group/cluster mass reconstruction techniques based
on galaxy properties. As the first step, we intentionally use a very
clean and straightforward set-up: a simple HOD galaxy mock catalogue
built upon a nearby-Universe light-cone. Later stages of this project
will involve more sophisticated mock galaxy catalogues using both more
advanced HOD models (Skibba et al. in preparation) and semi-analytic
modelling (\citealt{2006MNRAS.365...11C}). This paper sets out to
determine the simplest-case baseline by using a clean, well-defined
dataset with idealised substructure, sharp boundaries, spherically
symmetric haloes and a strong richness correlation. Given initial
estimates for the location of the structures, just how bad can it get?\\
\begin{figure}
 \centering
 \includegraphics[trim = 32mm 0mm 0mm 8mm, clip, width=0.58\textwidth]{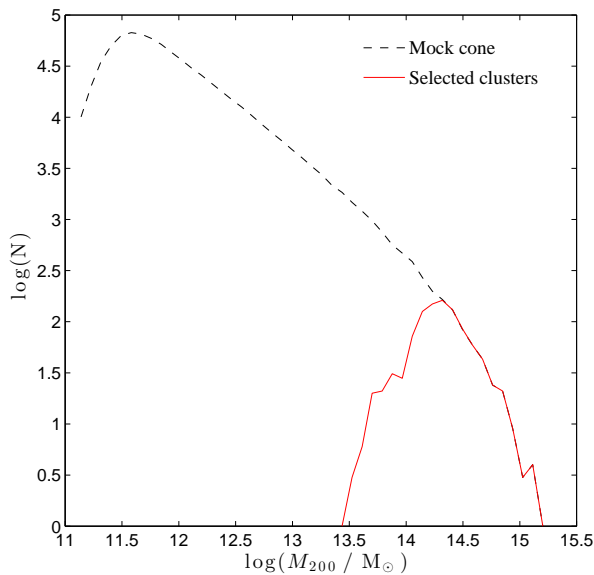}
 \caption{Cluster mass functions for the mock HOD light-cone (black dashed line) and the selected sample (red solid line). To deliver a sample with a sufficient number of
 cluster-sized haloes, the 1000 groups/clusters are selected by
 taking the 500 most massive, the next 300 richest and finally the groups/clusters with the most luminous
 brightest cluster galaxy (BCG) are taken to complete the sample.}
\label{fig:phase_1_mass_functions}
\end{figure}

\indent For Phase I, the dataset is the mock galaxy catalogue constructed in
\citet{2012MNRAS.419.2670M}. We briefly describe the catalogue here,
and we refer the reader to the above paper and to
\citet{2009MNRAS.392.1080S} for more details. We begin with the
Millennium Simulation (\citealt{2005Natur.435..629S}), which tracks
the evolution of $2160^{3}$ dark matter particles of mass $8.6 \times
10^{8}\,h^{-1} {\rm M_{\rm \odot}}$ from $z=127$ to $z=0$ within a
comoving box of side length $500\,h^{-1} {\rm Mpc}$, with a halo mass
resolution of $\sim5\times10^{10}h^{-1}\rm{M_\odot}$. The simulation
adopts a flat $\Lambda$CDM cosmology with the following parameters:
$\Omega_{\rm 0}=0.25$, $\Omega_{\rm \Lambda}=0.75$, $\sigma_{\rm
 8}=0.9$, $n=1$ and $h=0.73$. Collapsed haloes at $z=0$ with at least 20 particles are
identified with the \textsc{subfind} \citep{2001MNRAS.328..726S} FOF
group-finding algorithm, although consistent results are found with
other finders \citep{2011MNRAS.410.2617M, 2011MNRAS.415.2293K}.
The haloes are populated with galaxies whose luminosities and colours
follow the halo-model algorithm described in
\citet{2006MNRAS.369...68S} and \citet{2009MNRAS.392.1080S}, which is
constrained by the luminosity function, colour--magnitude
distribution, and luminosity- and colour-dependent clustering
\citep{2005ApJ...630....1Z} as observed in the Sloan Digital Sky
Survey \citep[SDSS;][]{2000AJ....120.1579Y}. An important assumption
in this HOD model is that all galaxy properties -- their numbers,
spatial distributions, velocities, luminosities, and colours -- are
determined by halo mass alone, again rendering the model as simple as
possible. We specify a minimum $r$-band luminosity for the galaxies
of $M_{r} = -19 + 5\log(h)$, to stay well above the resolution limit
of the Millennium Simulation.\footnote{The mass resolution of the simulation is sufficient that haloes that host galaxies as faint as $0.1~L_\ast$ ($M_r =−18+5\,\rm{log}(h)$) are typically resolved with more than $\sim$100 particles (\citealt{2005Natur.435..629S}), which corresponds to a stellar mass threshold of $M_\ast\sim10^{9.5}h^{-1}\rm{M_\odot}$.}\\
\begin{figure}
 \centering
 \includegraphics[trim = 35mm 0mm 0mm 0mm, clip, width=0.57\textwidth]{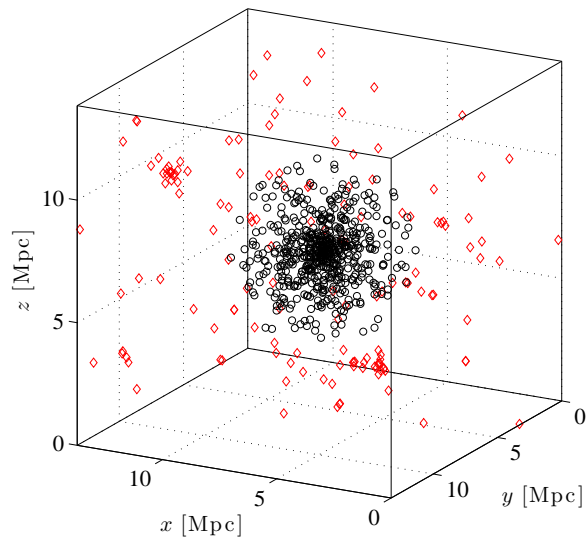}
 \caption{Real-space distribution of the HOD galaxies contained within
 the largest cluster (black circles) and surrounding galaxies
 contained within smaller haloes (red diamonds). The galaxy
 distribution is, by definition, spherical and lacking in
 substructure.}
\label{fig:phase_1_cluster_cube}
\end{figure}

\indent Haloes are assigned a `central' galaxy which has the same position and
velocity as the halo centre
(\citealt{2011MNRAS.410..417S}). `Satellite' galaxies are assumed to
be fainter than this object and follow an NFW density profile (\citealt{1997ApJ...490..493N}) that obeys the concentration--mass relation from
\citet{2008MNRAS.391.1940M}, with the population extending out to
$R_{\rm 200}$ (the radius that encloses a density 200 times the critical
density of the Universe), assuming isothermal, isotropic, Maxwellian velocity distributions.

\indent In the model of galaxy colours, central and satellite
galaxies have different colour--luminosity distributions. The central galaxy is usually the reddest galaxy in a given halo,
though satellites are redder than central galaxies at a given
luminosity (\citealt{2008MNRAS.387...79V};
\citealt{2009MNRAS.392.1467S}). Satellites are assumed to follow a
particular sequence on the colour--magnitude diagram, which approaches
the red sequence with increasing luminosity, consistent with what is
found in the SDSS group/cluster catalogues (\citealt{2009MNRAS.392.1467S}). 
Note that alternative approaches to modelling galaxy colours and colour-dependent clustering have recently appeared in the literature (\citealt{2013MNRAS.435.1313H}; \citealt{2013MNRAS.436.2286M}; \citealt{2013ApJS..208....1G}; \citealt{2013arXiv1312.1340P}).
 
\indent We also allow for the expected
scatter in the relation between host halo mass and central galaxy
luminosity \citep{2007ApJ...667..760Z}. The number of satellite
galaxies in a halo of given mass, $P(N_\mathrm{gal}|M)$, is
approximated well by a Poisson distribution
\citep{2004ApJ...609...35K}, with a mean HOD that increases
approximately linearly with mass, $\langle
N_\mathrm{sat}|M\rangle=[(M-M_0)/M_1^{ ' }]^\alpha$, hence we adopt this
distribution to populate the haloes\footnote{Note that the
 true relationship between richness and mass is determined by
 assumptions about the shape of the halo occupation distribution and
 its mean as a function of mass. The true number distribution at
 fixed mass is not Poisson, however, because of the group/cluster
 selection procedure (see Section~2).}. The value of $M_1^{ '
}/M_\mathrm{min}\approx17$ (where $M_\mathrm{min}$ is the mass
corresponding to the luminosity threshold: $M_r < −19 + 5\, \mathrm{log}\, h$), which determines the
critical mass above which haloes typically host at least one
satellite, is approximately independent of luminosity, and
$\alpha\approx1$ for most of this luminosity range
(\citealt{2011ApJ...736...59Z}). $M_0$ determines the shape of the satellite HOD at low halo masses and is typically smaller than $M_\mathrm{min}$. The HOD parameters are described in
detail in Appendix~A of \citet{2009MNRAS.392.1080S}. 

A galaxy's velocity is given by the sum of the velocity of its parent
halo plus an internal motion contribution within the halo. The internal motions are well approximated by a Maxwellian distribution (admittedly, $\Lambda$CDM haloes have more complex velocity distribution functions, see, e.g., \citealt{2001MNRAS.322..901S} and \citealt{2013arXiv1310.6756B}), with
velocities that are independent Gaussians in each of the three
Cartesian coordinates. The dispersion depends on halo mass and
radius through the scaling: 
\begin{equation}\sigma_\mathrm{\rm 200}^2=
GM_{\rm 200}/(2R_{\rm 200}).
\end{equation}
Note that this yields velocity dispersions that are $7\%$ greater than expected for NFW models with realistic anisotropic velocities (\citealt{2013MNRAS.429.3079M}). This overestimate of velocity dispersion and the assumption that it is independent of radius cause a violation of local dynamical equilibrium. 
In this phase of the project, we also neglect the effects of galaxy velocity bias 
(\citealt{2011MNRAS.410..417S}; \citealt{2013MNRAS.430.2638M};
\citealt{2013MNRAS.434.2606O}).

A $90 \times 90$ degree light-cone, $500\,h^{-1} {\rm Mpc}$ deep, is
constructed by taking a slice through the zero-redshift
simulation cone. To deliver a dataset with a sufficient number of
cluster-sized haloes, the 1000 groups/clusters are selected by
 taking the 500 most massive, the next 300 richest and finally the groups/clusters with the most luminous
brightest cluster galaxy (BCG) are taken to complete the sample. The
mass functions of the selected sample and the full light-cone are shown in Figure~\ref{fig:phase_1_mass_functions} via the solid red and black dashed line respectively.
An example of the underlying galaxy distribution inserted by the HOD
model is shown in Figure~\ref{fig:phase_1_cluster_cube}. Black circles
indicate the member galaxies for the largest cluster in the sample. By
construction, this spatial distribution is smooth and spherical,
lacking any imposed substructure, again to keep the test as simple as
possible. The red diamonds indicate galaxies in other haloes. As
Figure~\ref{fig:phase_1_cluster_cube} demonstrates, there are many
small haloes that have been populated with HOD galaxies that are not
part of our target list but which form the background galaxy
distribution for this initial phase of the project, to give the mass
measuring algorithms a simple contaminant to reject in their
analysis. Once a light-cone is generated, the internal velocity
dispersion of this large object will be added to the Hubble recession,
stretching it out along the line of sight, generating the usual
``Finger of God'' effect.

\indent In summary, the model for this simplified initial test generates data where clusters are spherically-symmetric, there is no internal
substructure, no galaxy velocity bias, no large-scale streaming motions, galaxies follow isotropic orbits and have effectively zero size so there is no blending of objects on the sky. 
\section{Mass Reconstruction Methods}
In this section, we present the halo mass reconstruction methods used
in this comparison project as listed in
Table~\ref{table:basic_method_characteristics}, which also summarises
some basic properties of each method. The following subsections
provide brief descriptions of each method, and are headed by an
identifying acronym used throughout this paper, as well as giving the names of the developers who participated in this project and the type of method involved.
\indent The two main steps performed by each method, the initial galaxy selection and the mass estimation, are separated into broad classes (which are specified in parentheses in the subsection titles). For the procedure of deducing the initial member galaxy sample, methods are categorised as either FOF (star), red sequence (diamond) or phase space (circle) -based. The mass estimation procedures are classed as either richness (magenta), phase space (black), radius (blue), abundance matching (green) or velocity dispersion (red) -based. Further details can be found in the paper
references that are provided in each description, and a more extensive
summary of the method characteristics is provided in the appendix.
\renewcommand{\tabcolsep}{0.4cm}
\begin{table*} 
 \caption{Summary of the participating cluster mass estimation
 methods. Listed is an acronym identifying the method, an
 indication of the scheme used to undertake member galaxy selection
 and an indication of the method used to convert this membership
 list to a mass estimate. Note that acronyms denoted with a star
 indicate that the method did not use our initial 1000 object
 target list but rather matched these locations at the end of their analysis. Please see Tables
 \ref{table:appendix_table_1} and \ref{table:appendix_table_2} in
 the appendix for more details on each method.} 
\begin{center} 
\begin{tabular}{l l l l}
\hline
\multicolumn{1}{c}{Method}&Initial Galaxy Selection&Mass Estimation&Reference \\ \hline
PCN&Phase space&Richness&Pearson et al. (in prep.)\\
PFN*&FOF&Richness&Pearson et al. (in prep.)\\
NUM&Phase space&Richness&Mamon et al. (in prep.)\\ 
ESC&Phase space& Phase space&{\citet{2013ApJ...768L..32G}}\\ 
MPO&Phase space& Phase space&{\citet{2013MNRAS.429.3079M}}\\ 
MP1& Phase space& Phase space&{\citet{2013MNRAS.429.3079M}}\\
RW& Phase space& Phase space&{\citet{2009MNRAS.399..812W}}\\
TAR*&FOF& Phase space&{\citet{2014arXiv1402.1350T}} \\
PCO& Phase space&Radius&Pearson et al. (in prep.)\\ 
PFO*&FOF& Radius& Pearson et al. (in prep.)\\
PCR& Phase space&Radius&Pearson et al. (in prep.)\\ 
PFR*&FOF&Radius&Pearson et al. (in prep.)\\
HBM*&FOF&Abundance matching&{\citet{2012MNRAS.423.1583M}}\\ 
MVM*&FOF&Abundance matching&{\citet{2012MNRAS.423.1583M}}\\
AS1&Red sequence&Velocity dispersion&{\citet{2013ApJ...772...47S}}\\
AS2&Red sequence&Velocity dispersion&{\citet{2013ApJ...772...47S}}\\
AvL& Phase space&Velocity dispersion&{\citet{2007MNRAS.379..867V}}\\ 
CLE& Phase space&Velocity dispersion&{\citet{2013MNRAS.429.3079M}}\\ 
CLN& Phase space&Velocity dispersion&{\citet{2013MNRAS.429.3079M}}\\
SG1& Phase space&Velocity dispersion&{\citet{2013ApJ...772...25S}}\\
SG2& Phase space&Velocity dispersion&{\citet{2013ApJ...772...25S}}\\
PCS& Phase space&Velocity dispersion&Pearson et al. (in prep.)\\ 
PFS*&FOF&Velocity dispersion&Pearson et al. (in prep.)\\ 
\hline 
\end{tabular}
\end{center}
\label{table:basic_method_characteristics}
\end{table*} 

\subsection{PCN (Pearson $\&$ Ponman, phase space, Richness)}
All PCx methods are based on a cylindrically selected galaxy sample. Starting with the halo positions, galaxies are initially selected from a $\rm 5\,Mpc$ radius cylinder about each halo with a depth of $\pm 1000\,{\rm ~km ~s^{-1}}$. The velocity depth is then iterated with a robust $3\,\sigma$ clipping using galaxies within $\rm 1\,Mpc$. To derive masses, the PCN method uses an aperture richness of each cluster as discussed in Pearson et al. (in preparation). Richness is defined as the number of galaxies above a threshold absolute magnitude within $\rm 1\,Mpc$, subtracting an interloper contribution estimated using galaxies in a background annulus of radii $\rm 3-5\,Mpc$. Mass is then estimated using a $M_{\rm 500}$-richness relation calibrated on a sample of clusters with SDSS galaxy data and X-ray estimates for $M_{\rm 500}$ from \citet{2009ApJ...693.1142S} and \citet{2010MNRAS.402...65S}. The estimated $M_{\rm 500}$ is converted to $M_{\rm 200}$ for this project using the mass-concentration relation of \citet{2008MNRAS.390L..64D}. We estimate statistical errors through bootstrap resampling the observed mass proxy and systematic errors by propagating errors on our calibration relation.

\subsection{PFN (Pearson $\&$ Ponman, FOF, Richness)}
All PFx methods are based on a FOF-selected cluster sample. We apply a FOF analysis using the scheme of \citet{2004MNRAS.348..866E} which utilizes the positions and velocities of galaxies. The linked clusters clusters are matched to the given cluster center positions. All linked galaxies are assumed to be cluster members, so we do not include any corrections for interloper contamination. For the PFN method, masses are derived based on the FOF richness of each cluster as discussed in Pearson et al. (in preparation), and calibrated against X-ray masses using the same sample as described for the PCN method, and is converted to $M_{\rm 200}$ in the same way. Statistical errors are estimated from Poisson errors propagated through the calibrated mass relation; systematic errors are derived from calibrating uncertainties as for PCN.

\subsection{NUM (Mamon, phase space, Richness)}
\indent The radius $R_{\rm 200}$ is estimated using the richness
measured in a rectangular area of projected phase space within $\rm 1
\,Mpc$ and $\rm 1333 \,km\,s^{-1}$ from the halo centre, with a linear
relation between $\log R_{\rm 200}$ and $\log N(1\,\rm
\,Mpc,1333\,km\,s^{-1})$ deduced from a robust linear fit to the
mock clusters analysed by CLE (see Sect. 19 below). The membership is deduced by selecting all
galaxies within $R_{\rm 200}$ and with velocities, relative to the
central halo, smaller (in absolute value) than $ 2.7 \,\sigma_{\rm
 los}(R)$ (computed from an NFW model, as in the CLE method). See
Mamon et al. (in preparation).

\subsection{ESC (Gifford $\&$ Miller, phase space)}
The caustic technique utilizes the radius-velocity phase space information of galaxies in clusters, as well as their dispersion, to estimate the escape velocity profile of the host haloes. The mass profile is inferred by
integrating the square of the escape velocity profile multiplied by a
parameter $\mathcal{F}_{\beta}$ which contains information on the
potential, density, and velocity anisotropy profiles of the halo along
with fundamental constants. $\mathcal{F}_{\beta}$ is treated as
approximately constant (see \citealt{1999MNRAS.309..610D} and
\citealt{2011MNRAS.412..800S}) with a value of 0.65 as found in
\citet{2013ApJ...773..116G}. Member galaxies are identified as those
within the escape velocity envelope in radius-velocity phase space and
within the estimated $R_{\rm 200}$ of the halo. This technique is
described in both \citet{2013ApJ...768L..32G} and
\citet{2013ApJ...773..116G}. 

\subsection{MPO (Mamon, phase space)}
Starting from the sample of members obtained with the CLN algorithm,
the virial radius, $R_{\rm 200}$, total mass scale radius, $R_{\rm
 \rho}$, red and blue galaxy population scale radii, $R_{\rm red}$
and $R_{\rm blue}$, and the velocity anisotropies at the virial radius
of these red and blue populations are computed using the Bayesian
MAMPOSSt method \citep{2013MNRAS.429.3079M}. This method jointly fits
the positions of the red and blue galaxies in projected phase
space. Here, it is assumed that the system is spherically symmetric
and that the total mass distribution follows the NFW model, while the
red and blue galaxy populations follow NFW models, each with its scale radius. The red and blue populations are assumed to have isotropic
orbits at the centre, but increasingly radial or tangential beyond
this (with different free outer anisotropies, but a transition scale fixed to be
the scale radius of the tracer). The 3D velocities are assumed to be
Gaussian at all radii. 

\subsection{MP1 (Mamon, phase space)}
MP1 is like MPO, but is colour-blind: a single tracer population is
assumed.

\subsection{RW (Wojtak, phase space)}
\indent In this method, the halo mass $M_{\rm 200}$ is derived from
the distribution of galaxies in phase space. It is assumed that the
galaxies follow a combination of a spherical NFW model (where number
follows mass) with a distribution function of energy and angular
momentum derived from $\Lambda$CDM haloes
(\citealt{2008MNRAS.388..815W}), forcing here the inner and outer
anisotropies to match those of $\Lambda$CDM haloes, and a constant
projected density background term that is kept as a free
parameter. See \citet{2009MNRAS.399..812W} for details. The membership
is determined by restricting to galaxies within $v_{\rm los} < \sqrt{-2
 \Phi(R)}$, where $R$ is the projected distance of the galaxy. 

\subsection{TAR (Tempel, FOF, phase space)}
TAR groups/clusters are based upon the conventional FOF group finding
algorithm, where the linking-length is calibrated based on the mean
distance to nearest galaxy in the plane of the sky. For the current
dataset $d_\perp=0.44\,h^{-1}\rm \,Mpc$ and $d_{||} = 440 \,\rm
km\,s^{-1}= 4.4\,h^{-1}\,\rm \,Mpc$ (assuming $d_{||} / d_\perp =
10$). More details of the group finding algorithm are explained in
\citet{2008A&A...479..927T,2010A&A...514A.102T} and
\citet{2012A&A...540A.106T}. The masses of groups/clusters are estimated by
applying the virial theorem to the sphere of radius $R_{\rm 200}$: 
\begin{equation}
	M = \frac{3\,R_{\rm G} \sigma_v^{2}}{G} = 7.0\times10^{12}\, \frac{R_{\rm G}}{\mathrm{\,Mpc}} \left(\frac{\sigma_v}{100\,\mathrm{km\,s}^{-1}}\right)^{2} M_{\sun} ,
\end{equation} 
where $\sigma_{v}$ is the 1D velocity dispersion. The gravitational
radius $R_{G}$ is estimated from the RMS projected radius. For that we
assume a NFW profile and find the theoretical relationship between
these two parameters. Since the concentration parameter of the NFW
profile depends on the halo mass (we use the mass-concentration
relation from \citealt{2008MNRAS.391.1940M}), we find the final mass
iteratively. See \citet{2014arXiv1402.1350T} for more details of
the method.

\subsection{PCO (Pearson $\&$ Ponman, phase space, Radius)} 
Using the galaxy membership of PCN, the galaxy overdensity profiles of clusters are modelled and fitted as described in Pearson et al. (in preparation). A projected NFW profile (\citealt{1996A&A...313..697B}) plus a uniform background term to allow for interloper contamination, is fitted to all galaxies within $\rm 5\,Mpc$. From the fitted NFW profile a radius $R_{500}$ is found, within which the cumulative number density is $500/\Omega_m$ times the mean cosmic number density of galaxies. This number density is estimated from the SDSS luminosity function of \citet{2003ApJ...592..819B} where galaxies are counted above a threshold luminosity of $M_r - \log h = -19$. The mass $M_{\rm 500}$ within $R_{500}$ is then deduced from $R_{500}$. These overdensity masses have been calibrated against the X-ray masses described under the PCN method, and as a result a linear scaling is applied to determine the final $M_{\rm 500}$ estimate, which is then extrapolated to $M_{\rm 200}$. Error analysis is as for PCN.

\subsection{PFO (Pearson $\&$ Ponman, FOF, Radius}
Using the linked galaxy membership of PFN, the galaxy overdensity profiles of clusters are modelled and fitted as described in Pearson et al. (in preparation). We fit a projected NFW (\citealt{1996A&A...313..697B}) profile, assuming that the linked galaxy membership is subject to no interloper contamination. $M_{\rm 200}$ is then derived from the fitted profile as for PCO.

\subsection{PCR (Pearson $\&$ Ponman, phase space, Radius)}
Using the galaxy membership within $\rm 1\,Mpc$, as derived for PCN, this method is based on the RMS radius of each cluster as discussed in Pearson et al (2013, in preparation). Note, however, that since we have no way of knowing which galaxies are interlopers, we are unable to make any statistical allowance for them (in contrast to the PCN method). As for PCN, we apply a relation calibrated on X-ray derived masses to estimate $M_{\rm 500}$, which is then extrapolated to $M_{\rm 200}$. Error analysis is as for PCN.

\subsection{PFR (Pearson $\&$ Ponman, FOF, Radius)}
The method is the same as PCR, except that it is applied to the FOF-selected galaxy membership described for PFN.

\subsection{HBM (Mu\~{n}oz-Cuartas, FOF, Abundance Matching)}
HBM is based upon an ellipsoidal FOF method with linking
lengths adapted according to the estimated halo mass. The linking
length along the line of sight is controlled by the (theoretical)
velocity dispersion of the halo. Cluster masses are determined by abundance matching between the cluster $r$-band luminosity function and the theoretical halo mass function of (\citealt{2006ApJ...646..881W}). The centre of the halo is set to the galaxy with the largest $r$-band luminosity. The
method is described in detail in \citet{2012MNRAS.423.1583M}. 

\subsection{MVM (M\"uller, FOF, Abundance Matching)}
MVM is the same as HBM with the difference that the center is assumed to lie at the center of stellar mass, while the virial theorem is used to compute $M_{\rm 200}$. The procedure is described in more detail in \citet{2012MNRAS.423.1583M}.

\subsection{AS1 (Saro, Red sequence, Velocity dispersion)}
AS1 was developed to study possible systematics affecting follow-up
dynamical mass estimation of high-redshift massive galaxy clusters. By
construction, it assumes that the centre of the cluster is known,
along with an initial estimate of $R_{\rm 200}$ from other
observables. It also assumes an intrinsic scatter of $\approx30\%$ in
mass at fixed velocity dispersion, mainly driven by the triaxial
properties of DM haloes. As the simulated clusters in this work are
spherical, this is largely overestimated. As the total estimated
errors on individual clusters mass could be larger than
$\approx 60\%$, it does not iterate to solve for $R_{\rm 200}$,
but is focussed more on obtaining an average unbiased mass estimation
for an ensemble of clusters. Since, for the purpose of this work, no
initial $R_{\rm 200}$ was given, it assumes a fiducial value of $\rm
1\,Mpc$ for all the mass range. Galaxies must lie within 0.1
magnitude in colour from a model given by \citet{2012ApJ...747...58S},
which has proven to be a good fit to the observational data and they must also lie within
$\rm 4000\, km\,s^{-1}$ from the cluster centre. A final clipping of
$\pm 3\, \sigma$ is then performed to remove interlopers and provide a
robust estimate of the velocity dispersion. A scaling relation,
provided in \citet{2013ApJ...772...47S}, is then used to convert the
velocity dispersion into $M_{\rm 200}$. The model is cosmologically
dependent at a background level and assumes a cosmology of $\Omega_{M} =
0.3$, $\Omega_{\Lambda}=0.7$ and $\rm h_{0} = 0.7$. More details of
this method are described in \citet{2013ApJ...772...47S}.

\subsection{AS2 (Saro, Red sequence, Velocity dispersion)}
AS2 follows the same procedure as AS1 but the estimated velocity
dispersion is corrected by taking into account the number of galaxies
as described by Equation 6 in \citet{2013ApJ...772...47S}. Note: the values of the constants $a$ and $b$ of the relation
$\rm{log}\, M_{\rm 200} = a + b\, \rm{log}\,\sigma_{v}$ employed by the ASx methods can be found in Table~\ref{table:a_b_constants}.

\subsection{AvL (von der Linden, phase space, Velocity dispersion)}
\indent This method is a relatively simple velocity dispersion
estimator, as used for SDSS clusters in \citet{2007MNRAS.379..867V}
and for EDisCS clusters in \citet{2008A&A...482..419M}. Galaxies are
iteratively selected to lie within $2.5\,\sigma_{v}$ and $ 0.8\,R_{\rm
 200}$ -- the latter is estimated from $\sigma_{v}$ by assuming the
virial theorem. These cuts are chosen to make the method relatively
insensitive to contamination from nearby structures; $R_{\rm 200}$ and
the final $\sigma_{v}$ are corrected for the expected bias from
sigma-clipping. Note: the values of the constants $a$ and $b$ of the relation $\rm{log}\, M_{\rm 200} = a + b\, \rm{log}\,\sigma_{v}$ employed by AvL can be found in Table~\ref{table:a_b_constants}.

\subsection{CLE (Mamon, phase space, Velocity dispersion)}
The initial membership is limited to $R<3 \rm \,Mpc$ and $|v| <
\rm{4000 \,km\,s^{-1}}$. A relative velocity gap technique
(\citealt{1993ApJ...404...38G}), with gapper coefficient $C=4$, is
initially applied to remove obvious interlopers (keeping the largest
subsample). The radius $R_{\rm 200}$ is first estimated from the
aperture velocity dispersion, where the measured value (using the
robust Median Absolute Deviation, see \citealp{1990AJ....100...32B}), is
matched to the aperture velocity dispersion at $R_{\rm 200}$. This is predicted
for a spherical single-component NFW model with concentration of $c =
4$, with the \citet{2005MNRAS.363..705M} velocity anisotropy profile
(with anisotropy radius equal to the scale radius of the NFW model, as
found for $\Lambda$CDM haloes by \citealt{2010A&A...520A..30M}). The
membership is recovered by selecting all galaxies within $R_{\rm 200}$
with velocities relative to the central one smaller (in absolute
value) than $2.7 \sigma_{\rm los}(R)$ (computed from an NFW model with
the same velocity anisotropy model, but now assuming a concentration
obtained from the $\Lambda$CDM concentration-mass relation of
\citealp{2008MNRAS.391.1940M}). The virial radius and membership are
iterated, now measuring the aperture velocity dispersion using the
unbiased standard deviation. The method is described in the appendix
of \citet{2013MNRAS.429.3079M}. Note: the values of the constants $a$ and $b$ of the relation
$\rm{log}\, M_{\rm 200} = a + b\, \rm{log}\,\sigma_{v}$ employed by CLE can be found in Table~\ref{table:a_b_constants}.

\subsection{CLN (Mamon, phase space, Velocity dispersion)}
CLN is similar to CLE, but now using the output of NUM as input. Note: the values of the constants $a$ and $b$ of the relation
$\rm{log}\, M_{\rm 200} = a + b\, \rm{log}\,\sigma_{v}$ employed by CLN can be found in Table~\ref{table:a_b_constants}.

\subsection{SG1 (Sif\'on, phase space, Velocity dispersion)}
Both SG1 and SG2 implement the shifting gapper of
\citet{1996ApJ...473..670F} and the velocity dispersion-mass relation
of \citet{Evrard:2008vo}. All galaxies within $4000\,\rm
km\,s^{-1}$ (rest-frame) of the cluster redshift are binned in
projected radial annuli, each of which has at least 15 galaxies and a
minimum width of $\rm 250\,kpc$. Galaxies within each bin are ordered
by the modulus of the velocity and a main body is defined by finding a
gap between two successive velocities of $500\,\rm km\,s^{-1}$ or
more. All galaxies within $\rm 1000\,km\,s^{-1}$ of this boundary are
considered halo members. The velocity dispersion is the bi-weight
estimate of scale (\citealt{1990AJ....100...32B}) of all members. From
this velocity dispersion, a mass, $M_{\rm 200}$, is estimated from the
velocity dispersion-mass relation of \citet{Evrard:2008vo}, and the
radius, $R_{\rm 200}$, is estimated from this mass. A new velocity
dispersion is computed using only members within $R_{\rm 200}$, and
this process is repeated until convergence (usually $\sim3$
iterations). The full description of the implementation
is in \citet{2013ApJ...772...25S}. 

\subsection{SG2 (Sif\'on, phase space, Velocity dispersion)}
SG2 is the same algorithm as SG1 but with different parameters for the
shifting gapper method: radial bins have a minimum width of $\rm 150
\,kpc$ and 10 galaxies; the main body boundary is $300\,\rm
km\,s^{-1}$ and all galaxies within $500\,\rm km\,s^{-1}$ of this
boundary are considered members. Consequently, SG1 and SG2 only differ in the membership selection. Note: the values of the constants $a$ and $b$ of the relation
$\rm{log}\, M_{\rm 200} = a + b\, \rm{log}\,\sigma_{v}$ employed by the SGx methods can be found in Table~\ref{table:a_b_constants}.

\subsection{PCS (Pearson $\&$ Ponman, phase space, Velocity dispersion)}
Using the galaxy membership within $\rm 1\,Mpc$, as for PCN, this method is based on the velocity dispersion of each cluster as discussed in Pearson et al. (in preparation). The velocity dispersion is determined using the Gapper estimator (\citealt{1990AJ....100...32B}). From the virial theorem, we expect $M \propto \sigma^3$. In practice, both the normalisation and power law index of the relation between mass and velocity dispersion has been calibrated to the X-ray derived masses, extrapolated to $M_{\rm 200}$, and errors estimated, as described for PCN. Note: the values of the constants $a$ and $b$ of the relation
$\rm{log}\, M_{\rm 200} = a + b\, \rm{log}\,\sigma_{v}$ employed by PCS can be found in Table~\ref{table:a_b_constants}.

\subsection{PFS (Pearson $\&$ Ponman, FOF, Velocity dispersion)}
This algorithm is identical to PCS, except that it uses the FOF-linked galaxy membership of PFN. Note: the values of the constants $a$ and $b$ of the relation
$\rm{log}\, M_{\rm 200} = a + b\, \rm{log}\,\sigma_{v}$ employed by PFS can be found in Table~\ref{table:a_b_constants}.
\renewcommand{\tabcolsep}{0.4cm}
\begin{table} 
 \caption{Values of the constants $a$ and $b$ of the relation
$\rm{log}\,( M_{\rm 200}/1x10^{14} M_{\odot}) = \it{a} + \it{b}\, \rm{log}\,(\sigma_{v}/1000kms^{-1})$ employed by methods that utilise the group/cluster velocity dispersion.
Please see Tables
 \ref{table:appendix_table_1} and \ref{table:appendix_table_2} in
 the appendix for more details on each method.} 
\begin{center} 
\begin{tabular}{c c c}
\hline 
\multicolumn{1}{c}{Method}&$a$&$b$ \\ \hline
AS1& 1.080&2.910\\
AS2& 1.080&2.910\\
AvL& 1.220& 3.000\\ 
CLE& 1.064&3.000 \\
CLN &1.064&3.000\\
SG1&1.034 &2.975\\
SG2& 1.034&2.975\\
PCS& 0.608&2.280\\ 
PFS*& 0.797&2.750\\ 
\hline 
\end{tabular}
\end{center}
\label{table:a_b_constants}
\end{table} 
\begin{figure*}
 \centering
 \includegraphics[trim = 0mm 15mm 0mm 25mm, clip, width=1.0\textwidth]
 {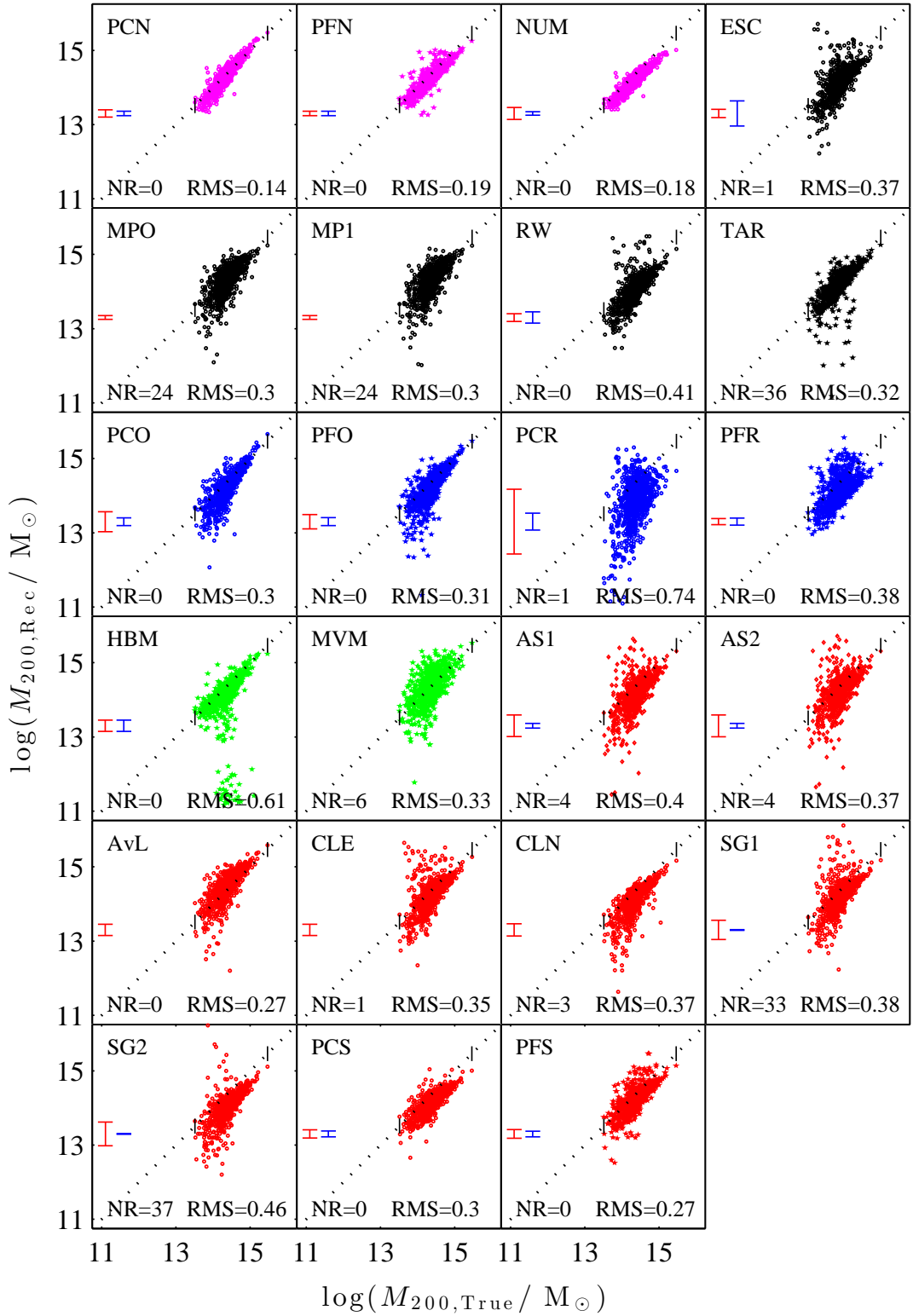} 
 \caption{Recovered versus true mass when
 the group/cluster membership is not known. The black dotted line represents
 the 1:1 relation. `NR' in the legend represents groups/clusters
 that are not recovered because they are found to have very low
 ($\rm < 10^{10} \,M_{\odot}$) or zero mass. The black ticks that
 lie across the 1:1 relation represent the minimum and maximum
 `true' halo $M_{\rm 200}$. The vertical red bar represents the mean
 statistical error delivered by methods and the vertical blue bar
 represents the mean systematic error delivered by methods.} 
\label{fig:phase_1_mass_scatter_subplots}
\end{figure*}
\begin{figure*}
 \centering
 \includegraphics[trim = 30mm 25mm 0mm 23mm, clip, width=1.25\textwidth]{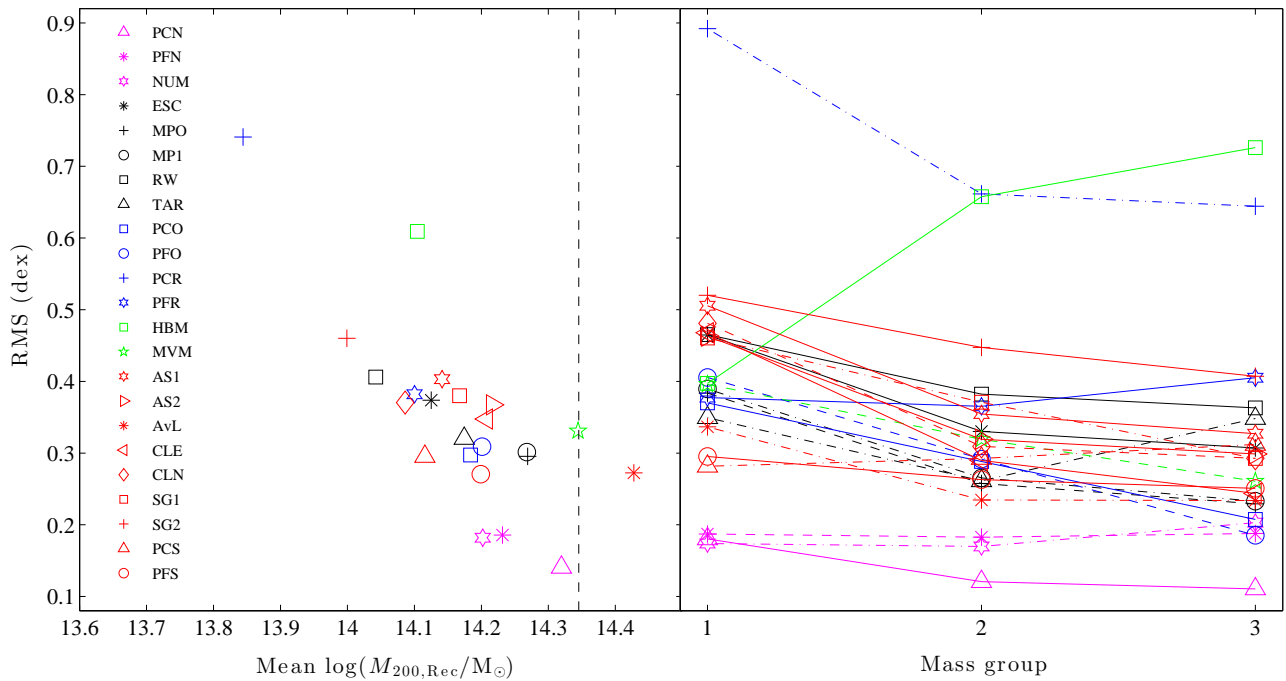}
 \caption{Left hand side: RMS error on log mass versus mean recovered mass (in dex) when the true galaxy membership is not known. The dashed black line identifies where the mean of the true mass distribution lies. Right hand side: RMS error on log mass versus mean recovered mass (in dex) for three mass bins when the true galaxy membership is not known. Mass groups 1, 2 and 3 represent clusters with `true' $M_{\rm 200}$ within the ranges log$(M_{\rm 200}) \leq 14.25$, $ 14.25 < \mathrm{log}(M_{\rm 200}) \leq
14.45$ and $14.45 < \ \mathrm{log}(M_{\rm 200})$, respectively.}
\label{fig:phase_1_rms_combined}
\end{figure*}
\begin{figure}
 \centering
 \includegraphics[trim = 39mm 0mm 0mm 10mm, clip, width=0.57\textwidth]{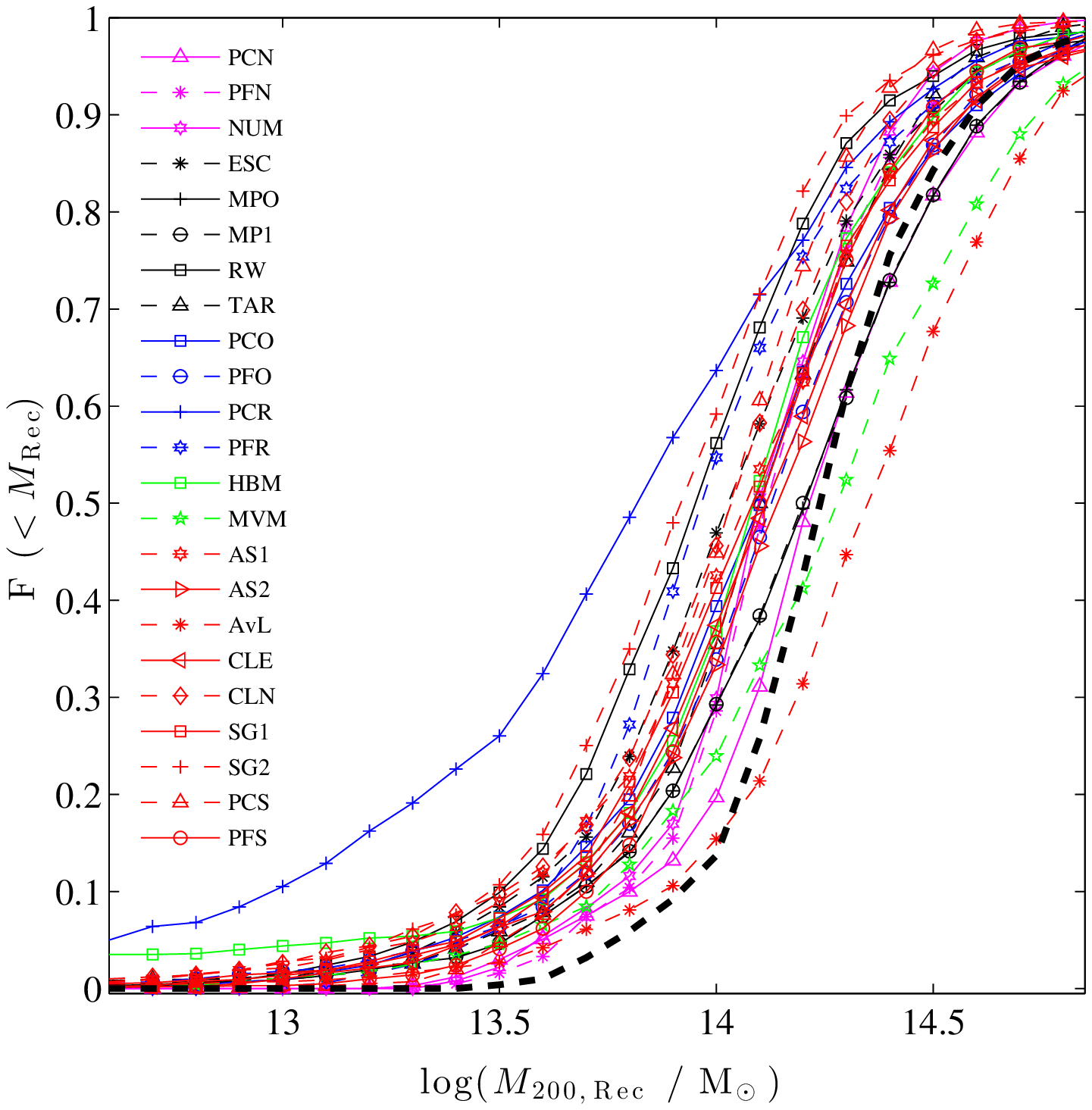}
 \caption{The cumulative distribution functions (CDF) of recovered halo mass delivered by the 23 methods when the group/cluster membership is not known. The true $M_{\rm 200}$ CDF is shown via the thick black dashed line. It is clear that all methods are recovering lower halo masses than that of the 'true' mass. A two-sample Kolmogorov-Smirnov (KS) test also demonstrates that the recovered mass distributions are not statistically similar to the true mass distribution (delivering $p$-values for all methods $p \leq 0.01$).}
\label{fig:phase_1_CDF}
\end{figure}
\section{Results: Cluster Mass Comparison}
In this section, we present results comparing the recovered group/cluster masses of the different reconstruction methods to the `true' masses in
the catalogues. We first make comparisons when the galaxy members are
selected by the algorithms, before examining the simpler case
where actual galaxy membership is specified (i.e., the case in which membership is known \textit{a priori}). The differences between
these results will allow us to distinguish between uncertainties due
to the mass estimates and uncertainties due to the identification of
group and cluster members. We note that supplying the `true' galaxy
membership does not necessarily guarantee an improvement, nor should it
 be expected: the methods have at most one or two free
parameters, and are not generally tuned to each specific dataset but
are rather run using their default settings. As such, if the default
setting assumes a more restrictive or extensive galaxy membership, the
calculation of mass from galaxy members selected in a different way is not guaranteed to be
reliable. However, the level of scatter will indicate
whether the calculation of mass from galaxy members selected in a different way is a serious issue.

\indent Figure~\ref{fig:phase_1_mass_scatter_subplots} shows the recovered
mass versus input mass when the group/cluster membership is \textit{not} given
in advance. The colour scheme separates the methods into the broad
classifications introduced in the last section and are classed as either
richness (magenta), phase space (black), radius (blue), abundance matching (green) or velocity dispersion (red) -based. Methods where membership is based on an FOF linking method have star-shaped markers, red sequence-based methods have diamond markers and phase space-based methods have circle markers. If returned,
the vertical bars at the left hand side of each panel indicate the
statistical (red) and systematic (blue) errors estimated internally by
the mass reconstruction methods themselves without reference to this
plot.

\indent Encouragingly, we see a correlation across the input mass range $13<\mathrm{log}(M_{\rm 200}/\rm{M_\odot})<15$. There is generally good
agreement, at least for the inferred masses of massive galaxy
clusters. Nonetheless, one can see substantial scatter, especially at
masses of $\mathrm{log}~(M_{\rm 200}/\rm{M_\odot})<14$, typically associated
with groups. Although for some methods or mass regimes the masses may
be slightly overestimated, the masses of groups and poor clusters
appear to be more often underestimated, except for the methods based
upon richness (PCN, PFN and NUM), as well as HBM. These biases are
also apparent in Figure~\ref{fig:phase_1_mass_residual}, which shows the residual recovered mass in dex via log($M_{\rm 200, Rec} /  M_{\rm 200, True}$). In some cases, these
masses may be underestimated by more than an order of
magnitude. Phase space methods MPO, MP1, TAR and velocity dispersion methods SGx fail to recover masses for some groups/clusters,
but the number of such cases amounts to a very low fraction of the
sample. The PCR method fails to recover reliable masses. This
method uses the RMS radius of the galaxy distribution extracted
within a $\rm 1\,Mpc$ aperture (and velocity range). However, this parameter is
inflated by the presence of interlopers (which can removed statistically
when calculating richness, for example) and is reduced by the imposed
$\rm 1\,Mpc$ aperture. It is noticeable that the PFR method, which is also
based on RMS radius, but is less affected by interlopers and has no
restrictive aperture imposed, performs significantly better.

\indent In the left hand side of Figure~\ref{fig:phase_1_rms_combined}, we quantify the error in the
estimated masses by calculating the RMS of the
difference between the recovered mass and the input mass (in $\rm
dex$) and we display this versus the mean of the recovered mass
distribution. The black dashed vertical line identifies the mean of
the true mass distribution. It is clear that the majority of methods,
with the exception of AvL and MVM, are systematically biased to lower halo
masses. For the majority of the methods, the RMS error on $M_{\rm 200}$ is
of the order of $\rm 0.3\,dex$ (i.e. a factor of 2). Richness-based
methods NUM, PFN and PCN produce the lowest RMS values indicating
lower scatter. Both PCR and HBM are outliers, delivering substantially
higher RMS values than other methods. For HBM, this higher scatter is most
likely due to the large tail of groups/clusters recovered with low
masses as seen in Figure~\ref{fig:phase_1_mass_scatter_subplots}. As
we will see below, these lower masses are not seen when the
galaxy membership is defined, and they seem to be due to the galaxy
selection algorithm returning very few galaxy members (most likely due
to a mis-matching cluster centres when HBM performs the initial step
of cluster finding).

\indent We present the cumulative distribution functions (CDFs) of
the recovered halo masses in Figure~\ref{fig:phase_1_CDF}. The CDFs illustrate the mass range over which a given method tends to under/overestimate halo masses: most methods are biased low over the entire mass range, while a few methods (AvL and MVM) are biased high for massive clusters $\mathrm{log}~(M_{\rm 200}/\rm{M_\odot}) \geq 14.2$. While only $\sim10\%$ of the input groups/clusters have a mass of $\mathrm{log}~(M_{\rm 200}/\rm{M_\odot}) \leq 14$, some methods assign $\sim65\%$ of the population a mass of $\mathrm{log}~(M_{\rm 200}/\rm{M_\odot}) \leq 14$, highlighting further that the majority of methods are recovering lower group/cluster masses than one would expect.
Those derived using the 23 methods reveal that none of the algorithms
return a measured mass distribution consistent with the input data
(with p-values for all methods $\leq 0.01$). These recovered mass distributions for all 23 methods can be seen in Figure~\ref{fig:phase_1_mass_hists} in the appendix.

\indent To quantitatively compare how well the different methods reconstruct
group/cluster mass, we calculate the difference between the recovered
mass and the true group/cluster $M_{\rm 200}$ via $| \rm{log}\,(M_{\rm 200,
 True}/M_{\rm 200, Rec.})|$. The mean of these values is taken to
calculate the mean deviation along with the dispersion of the deviations. The RMS of these values is used to rank
the 23 methods as shown in the final column of Table
\ref{table:mass_deviation_table}. The method producing the lowest RMS
is given a ranking of 1 and the method producing the highest mean
deviation is given a ranking of 23. The RMS ranking where the average bias of a given method has been subtracted is also given in the second to last column ($\rm Rank_{\rm \sigma}$). Additionally, we separate the
groups/clusters into three `true' $M_{\rm 200}$ mass bins: log$(M_{\rm 200}) \leq 14.25$, $ 14.25 < \mathrm{log}(M_{\rm 200}) \leq
14.45$ and $14.45 < \ \mathrm{log}(M_{\rm 200})$, to explore
whether the mean deviation values for each method are consistent
across all masses. As seen earlier, we find that the majority of
methods have a higher mean deviation for groups/clusters in the lowest
group/cluster mass bin, this is highlighted in the right hand side of Figure~\ref{fig:phase_1_rms_combined} where the RMS of the difference between the recovered and input masses (in $\rm dex$) is shown for the three mass groups. It is also clear the three richness-based
methods recover the group/cluster masses well. The majority of the
remaining methods are very similar, with typical mass estimation
errors of a factor of 2 to 3.

\begin{figure*} 
 \centering
 \includegraphics[trim = -5mm 15mm 0mm 25mm, clip, width=1.0\textwidth]{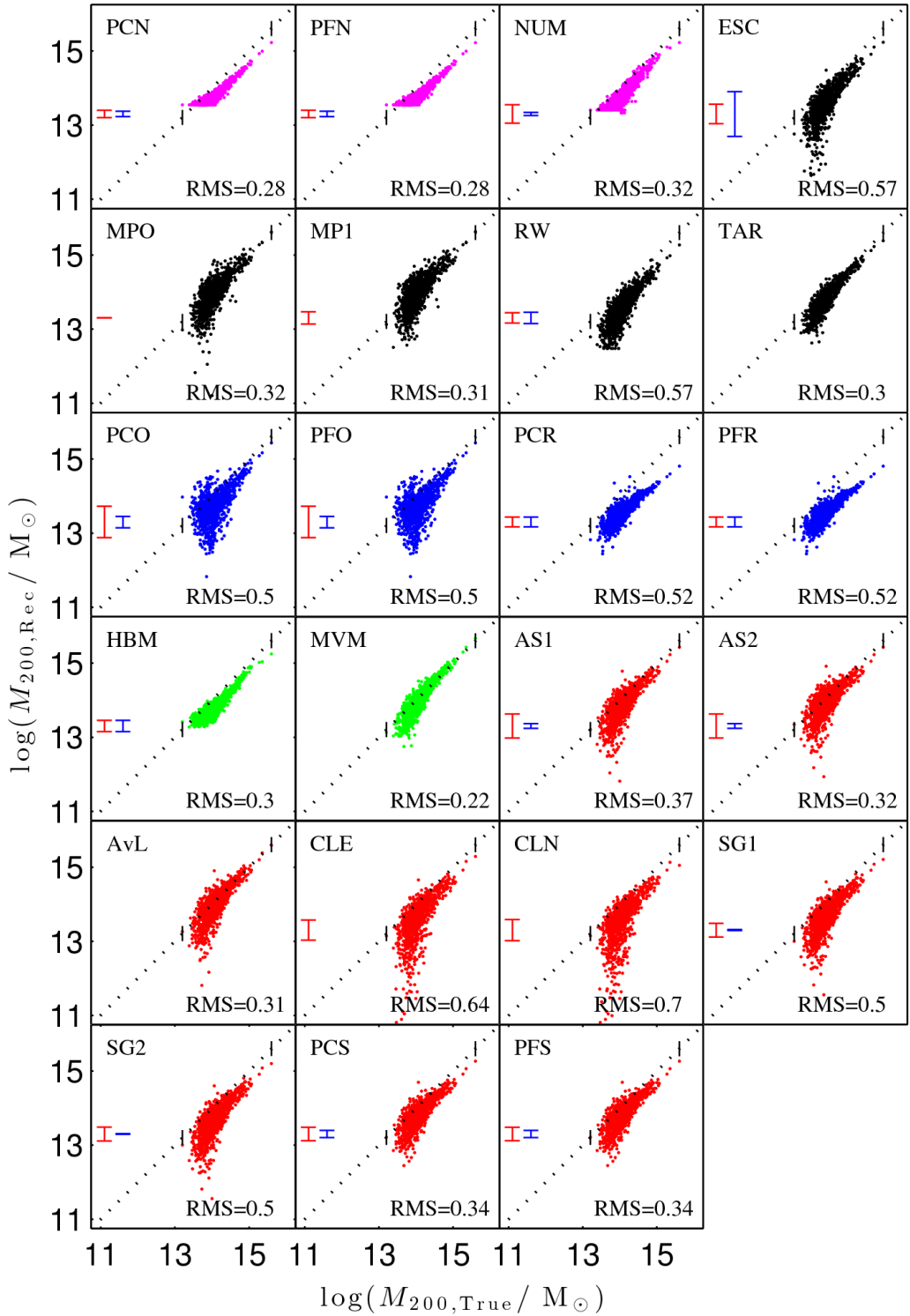}
 \caption{The true cluster mass versus recovered mass when the group/cluster membership is known. The black dotted line
 represents the 1:1 relation. `NR' in the legend represents
 groups/clusters that are not recovered because they are found to
 have very low ($\rm < 10^{10} M_{\rm \odot}$) or zero mass. The
 black ticks that lie across the 1:1 relation represent the minimum
 and maximum input group/cluster $M_{\rm 200}$. The vertical red
 bar represents the mean statistical error delivered by methods and
 the vertical blue bar represents the mean systematic error
 delivered by methods.} 
\label{fig:phase_1_mass_scatter_true_membership}
\end{figure*}

\begin{figure}
 \centering
 \includegraphics[trim = 35mm 0mm 0mm 5mm, clip, width=0.6\textwidth]{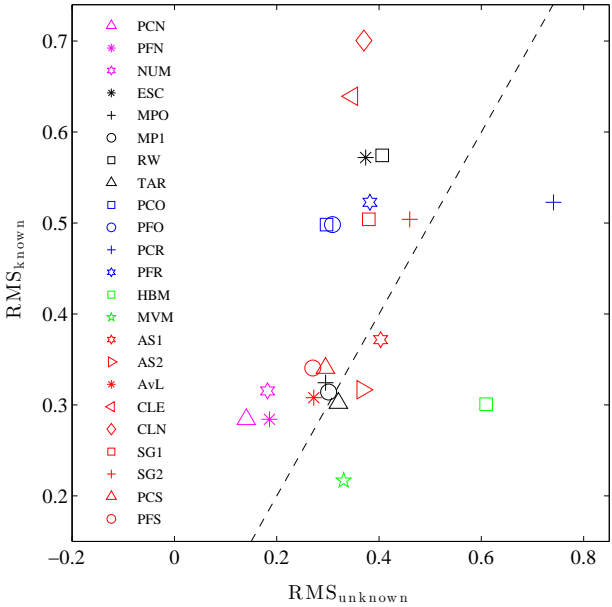}
 \caption{RMS difference in recovered versus true log mass when the membership is known versus that when the membership is not known (in $\rm dex$). Groups/clusters that are not
 recovered by the methods are excluded in this calculation. The
 black dashed line represents a 1:1 relation. The majority of methods have a
 higher RMS when estimating mass using the true membership.} 
\label{fig:phase_1_unknown_vs_known_rms}
\end{figure}

\indent In Figure~\ref{fig:phase_1_mass_scatter_true_membership}, we
show the recovered mass versus the input mass when each
group/cluster's galaxy membership is specified in advance. 
Note that though this group/cluster catalogue is derived from the same light-cone, the sample is not exactly the same as those in the previous figures. In order to maintain a blind set-up for comparison, new clusters from the light-cone were added into the sample. As a result, this sample has, on average, poorer groups/clusters.
One can see qualitatively similar correlations as in the previous results, 
but detailed comparisons indicate interesting differences between them. 
We immediately see that when the methods are not allowed to
restrict the galaxy membership according to their usual scheme, for the majority of methods, the
recovered masses are poorer (AvL, SG1, SG2, PFS, PCS, PFN, PCN, NUM, CLE, CLN, ESC, PFR, HBM, PCO, PCR, RW, MPO and MP1). This phenomenon is highlighted
in Figure~\ref{fig:phase_1_unknown_vs_known_rms}, where the RMS of the
difference between the recovered mass and true group/cluster mass is
calculated for both cases of unknown and known membership (in $\rm
dex$).
Note that groups/clusters that are not recovered by the methods are
excluded in this calculation.

\indent The majority of methods show a higher RMS when
estimating mass using the true membership as opposed to selecting
their own member galaxies. Furthermore, the widths of the
distributions (as shown in
Figure~\ref{fig:phase_1_mass_hists_truemembership} in the appendix) are not
significantly decreased; indeed in some cases they are increased. Moreover, the tail of underestimated group masses is more
pronounced. Some of the methods (e.g., ASx, SGx, CLE, ESC and RW) exhibit occasional large mass overestimates when they select
their own membership in a manner that does not occur when the galaxy
membership is fixed. These overestimates are not driven by nearby
large objects, as the number of member galaxies in these objects is
recovered approximately correctly. It seems likely that the dynamical
mass estimator is failing due to the influence of a small number of
interloper galaxies.

\indent Those methods which perform significantly better when provided
with the true galaxy membership (HBM, MVM and the two ASx methods)
have been calibrated using the true membership of haloes derived from
cosmological simulations, so it is natural that they should
perform best when provided with a set of galaxies which is
not contaminated by interlopers. In contrast, the PCx and PFx methods,
for example, have been calibrated using galaxy samples which
contain interlopers, and so one would expect their results to
be biased when given only the true group/cluster members.
\begin{table*} 
\caption{The mean, dispersion, RMS and ranking
 of $|$log$(M_{\rm 200, True}/M_{\rm 200, Rec.})|$ for three `true' mass bins:
 log$(M_{\rm 200}) \leq 14.25$, $ 14.25 < \mathrm{log}(M_{\rm
 200}) \leq 14.45$ and $14.45 < \ \mathrm{log}(M_{\rm 200})$. The bins are chosen so that there are roughly equal
 numbers of clusters in each mass bin. Here `1' represents the method
 with the lowest RMS and `23' represents the method with the highest
 RMS. Groups/clusters that are not recovered by the methods are
 excluded in this calculation. The overall RMS ranking calculated for
 groups/clusters of all masses where the average bias of a given method has been subtracted, $\rm Rank_{\rm \sigma}$, is given in the second to last column column. The overall RMS ranking calculated without bias subtraction is given in the final column.} 
\begin{center}
\begin{tabular}{p{0.3cm}p{0.25cm}p{0.25cm}p{0.25cm}P{0.25cm}p{0.25cm}p{0.25cm}p{0.25cm}P{0.25cm}p{0.25cm}p{0.25cm}p{0.25cm}P{0.25cm} P{0.45cm}P{0.45cm}}
\hline
\multicolumn{1}{c}{Method}&
 \multicolumn{4}{c}{$\mathrm{log} (M_{\rm 200}) \leq 14.25$}&\multicolumn{4}{c}{$14.25<\mathrm{log} (M_{\rm 200}) \leq 14.45$}
&\multicolumn{4}{c}{ $\mathrm{log} (M_{\rm 200}) \geq 14.45$}& \multicolumn{2}{c}{All masses}
 \\[1.0ex]
&Mean&$\;\,\sigma$&RMS&Rank&Mean&$\;\,\sigma$&RMS&Rank&Mean&$\;\,\sigma$&RMS&Rank&$\rm Rank_{\rm \sigma}$&$\text{\rm Rank}$\\[1.0ex]
 \hline 
PCN&0.14&0.12&0.18& 2&0.10&0.07&0.12& 1&0.08&0.07&0.11& 1& 2& 1\\
PFN&0.14&0.13&0.19& 3&0.15&0.11&0.18& 3&0.17&0.08&0.19& 3& 3& 3\\
NUM&0.14&0.11&0.17& 1&0.15&0.09&0.17& 2&0.18&0.09&0.20& 4& 1& 2\\
ESC&0.36&0.30&0.46&17&0.28&0.18&0.33&16&0.27&0.15&0.31&15&16&16\\
MPO&0.28&0.26&0.38&10&0.20&0.17&0.26& 5&0.17&0.15&0.23& 6&12& 7\\
MP1&0.28&0.27&0.39&11&0.20&0.17&0.27& 8&0.17&0.16&0.23& 7&13& 9\\
RW &0.38&0.26&0.47&18&0.34&0.18&0.38&20&0.33&0.15&0.36&19& 9&20\\
TAR&0.23&0.26&0.35& 7&0.19&0.18&0.26& 6&0.21&0.28&0.35&18&10&11\\
PCO&0.28&0.24&0.37& 8&0.23&0.18&0.29& 9&0.15&0.15&0.21& 5& 6& 8\\
PFO&0.26&0.31&0.41&14&0.21&0.21&0.29&11&0.13&0.13&0.19& 2&11&10\\
PCR&0.68&0.57&0.89&23&0.53&0.40&0.66&23&0.55&0.34&0.64&22&22&23\\
PFR&0.32&0.20&0.38& 9&0.33&0.16&0.37&18&0.37&0.17&0.41&20&14&18\\
HBM&0.19&0.35&0.40&13&0.32&0.58&0.66&22&0.31&0.66&0.73&23&23&22\\
MVM&0.29&0.27&0.40&12&0.24&0.21&0.32&14&0.21&0.16&0.26&11&19&12\\
AS1&0.37&0.35&0.51&21&0.27&0.23&0.35&17&0.25&0.21&0.33&17&21&19\\
AS2&0.32&0.33&0.46&16&0.23&0.22&0.32&15&0.22&0.20&0.30&14&20&14\\
AvL&0.25&0.22&0.34& 6&0.19&0.14&0.23& 4&0.17&0.16&0.23& 8& 7& 5\\
CLE&0.34&0.32&0.47&19&0.23&0.18&0.29&10&0.20&0.14&0.24& 9&17&13\\
CLN&0.34&0.34&0.48&20&0.25&0.18&0.31&13&0.23&0.18&0.29&13& 8&15\\
SG1&0.35&0.29&0.46&15&0.29&0.24&0.37&19&0.24&0.16&0.29&12&18&17\\
SG2&0.40&0.33&0.52&22&0.39&0.22&0.45&21&0.37&0.17&0.41&21&15&21\\
PCS&0.23&0.16&0.28& 4&0.25&0.15&0.29&12&0.28&0.13&0.31&16& 4& 6\\
PFS&0.23&0.19&0.30& 5&0.21&0.16&0.26& 7&0.22&0.12&0.25&10& 5& 4\\
\hline
\end{tabular}
\end{center}
\label{table:mass_deviation_table}
\end{table*}
\section{Results: Cluster membership} 
\begin{figure*}
 \centering
 \includegraphics[trim = -5mm 14mm 0mm 25mm, clip, width=1.02\textwidth]{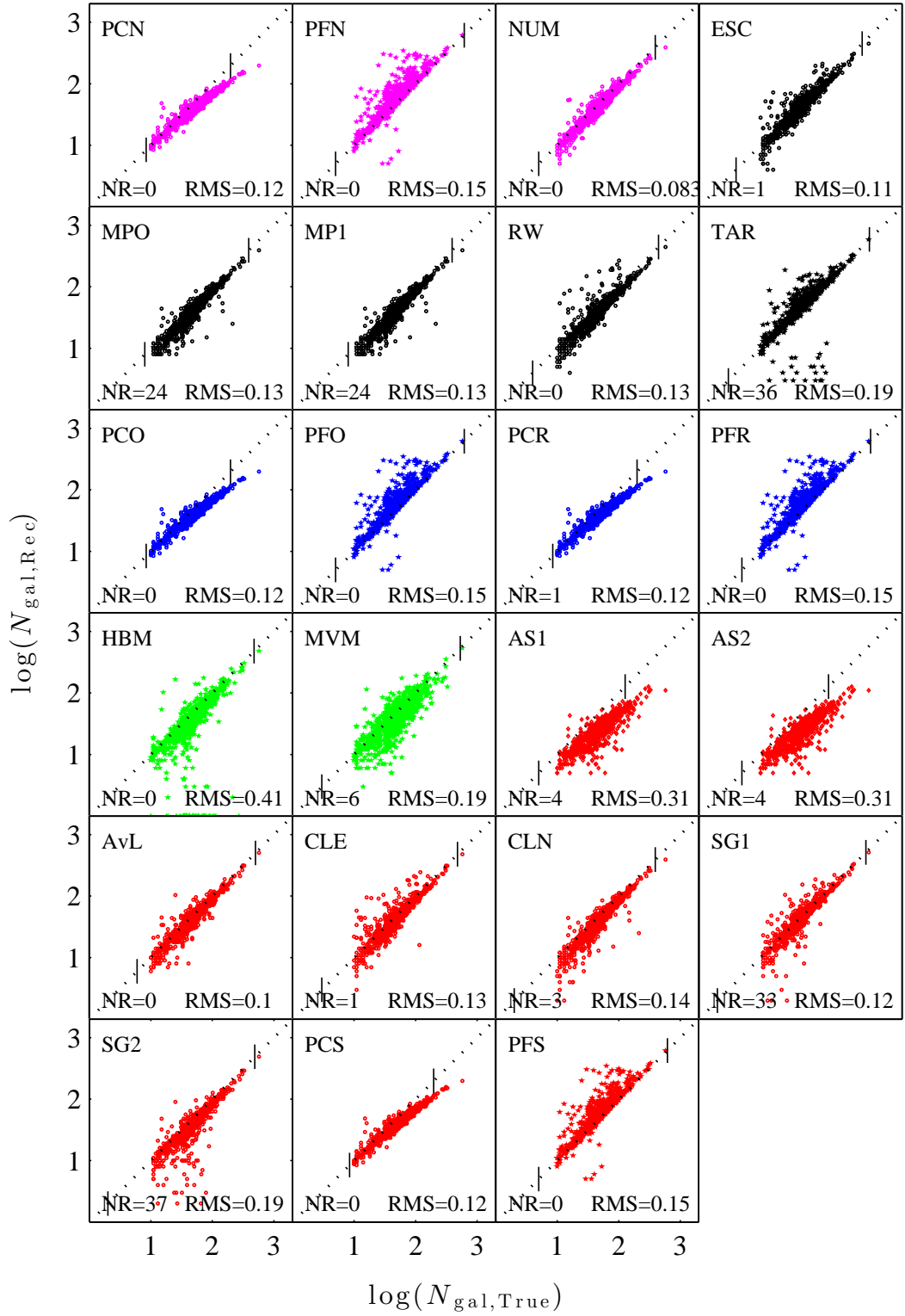}
 \caption{Recovered number of galaxies associated with each group/cluster
 versus the true number of galaxies when the group/cluster membership is not known. The
 black dotted line represents the 1:1 relation and the black ticks
 represent the true minimum and maximum number of galaxies
 associated with the input groups/clusters. `NR' in the legend represents groups/clusters that are not recovered because
 they are found to have very low ($\rm < 10^{10} \rm M_{\odot}$) or zero mass.} 
\label{fig:phase_1_ngaltrue_nalcat}
\end{figure*}

We now examine the galaxy cluster membership delivered by the various
methods and compare the richnesses of the recovered systems. Figure~\ref{fig:phase_1_ngaltrue_nalcat} presents the
richness of the recovered groups and
clusters compared to the number of member galaxies in the source
catalogue.

\begin{figure*}
 \centering
 \includegraphics[trim = 0mm 15mm 0mm 26mm, clip, width=1.0\textwidth]{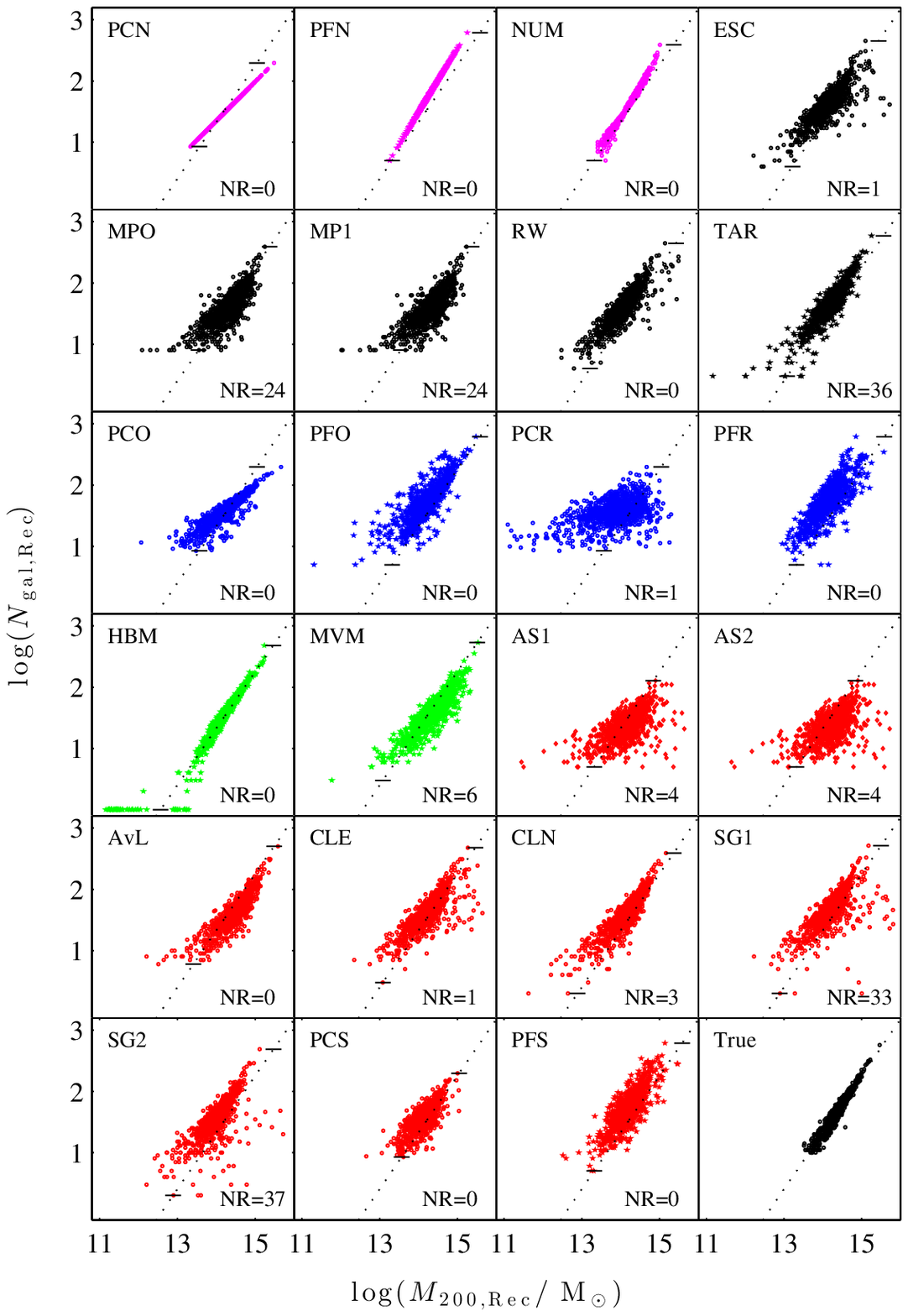}
 \caption{Recovered richness versus recovered mass for each halo, when the group/cluster membership is not known. The black dotted line represents the
 true mass versus the true number of galaxies associated with each
 halo and the black ticks represent the true minimum and maximum
 number of galaxies associated with the input
 groups/clusters. `NR' in the legend represents groups/clusters that
 are not recovered because they are found to have very low ($\rm <
 10^{10} \rm M_{\odot}$) or zero mass. The bottom right panel
 displays the input HOD mass-richness distribution.} 
\label{fig:phase_1_mass_ngal}
\end{figure*}

In general, the recovery of galaxy membership is very good
and we find tighter relations with somewhat lower levels of scatter in
comparison to the mass recovery results (also highlighted in Figure~\ref{fig:phase_1_ngaltrue_nalcat_residual} in the appendix). Certain methods tend to miss
members of massive clusters, such as both the AS and PC
approaches. This deficit is intrinsic to these methods, in that they
are deliberately conservative in their membership selection, focussing
on the very central regions of each object; the bias that this
selection introduces in the recovered mass is then calibrated out of the estimator. Other methods,
such as CLE, ESC, PFx, RW, and SGx, are more contaminated by interlopers and 
consequently have richness estimates that are biased
high. As mentioned earlier, this plot also illustrates that the high
recovered mass up-scattered clusters seen in the ASx, ESC and SGx
methods in Figure~\ref{fig:phase_1_mass_scatter_subplots} do not seem
to be due to line-of-sight contamination by higher mass objects as
very few objects have spuriously high numbers of recovered galaxies
for these methods.\\
\indent Most methods have distributions similar in shape to the true sample (as shown in Figure~\ref{fig:phase_1_ngal_hists} in the appendix)
although some distributions are, as noted above, offset either to high
values due to interloper inclusion or low values due to conservative
membership criteria.
Finally, Figure~\ref{fig:phase_1_mass_ngal} compares the recovered
richnesses as a function of recovered mass for the different
methods. The three richness-based methods, PFN, PCN, and NUM, have, as
expected, very tight relations. For both the input catalogue and these methods,
there is a close relationship between richness and cluster mass which
may not hold in the real universe. This strong correlation is a
consequence of the simple model that we have adopted for this initial
part of the comparison project. HBM, a velocity-dispersion-based
method, also has a particularly narrow distribution combined with some
catastrophic failures due to the mis-matching of a small number of
groups/clusters. In contrast, many other methods have more scatter in
both recovered number and recovered mass.
Moreover, the slope of the $N(M)$ relation is not well recovered for some methods (e.g., ASx, PCx) because of the variation in completeness as a function of mass seen in Figure~\ref{fig:phase_1_ngaltrue_nalcat} for these methods: although they recover the halo masses to a similar level of accuracy as the other methods, they should not be used for reliable member galaxy determination.
\section{Discussion}
\label{sec:Discussion}
The initial set-up used for this project was kept deliberately simple. 
We began with a simulated dark matter halo catalogue, and a model that inserts
galaxies via smooth, spherically-symmetric NFW distributions centred
at the centre of the dark matter potential well and scaled by the mass of the halo. 
Within the $z=0$ snapshot, haloes of mass above $\rm
10^{11.5}M_\odot$ (Figure~\ref{fig:phase_1_mass_functions}) are
populated and a light-cone is then drawn through this distribution to
create the ``observations'' used for this test. Once this baseline
study has quantified and minimised the uncertainties
intrinsic in mass estimation, we will move on to a more sophisticated
cluster model to identify the additional levels of uncertainty that such
complexity introduces. Due to the simplicity of the model used for this
initial phase and the use of a single cosmology, we cannot comment on
the absolute calibration of each model, other than noting that the
values have been calibrated to at least approximate reality. The main
focus of this paper is to quantify the underlying
scatter inherent in cluster mass estimation techniques that use the
positions, velocities, and colours of galaxies.

\indent There are three general stages involved in galaxy-based cluster mass
estimation. The first stage is the identification of a group/cluster
overdensity, the second is the selection of galaxies deemed to be
group/cluster members, and the third is the estimation of cluster
properties based on this membership. These steps are not, in practice,
independent from each other. For instance, a cluster mass estimation
method based on dynamical properties might be very sensitive to
contamination by unrelated field galaxies. As such, it is perhaps
better in such a method to be very conservative with the membership
selection at the expense of completeness and then recalibrate the mass
estimate based on this incomplete galaxy sample. Conversely, a method
based on the volume covered might not be sensitive to interlopers but
highly reliant on obtaining a nearly complete galaxy sample.

\indent Following the philosophy of this study of making things as simple as
possible, and to aid the inter-comparison of the results of different
methods, we supplied the participants with a list of initial centres
(i.e. the first stage of this process) about which to look for
structures. We further note that not all methods taking part in this
study include this step. The centres of the group/cluster sample
correspond with the location of the brightest cluster galaxy in all
cases and are the ``true'' location of the halo centre in the DM
simulation (the HOD model used places the brightest galaxy at the
location of the most bound DM particle in the halo). Some methods
(indicated by an asterisk in Table~1) chose not to use this
information, and instead used the full galaxy catalogue detecting
initial centres themselves. After calculating the properties of the
identified groups/clusters, these methods then matched to our supplied
coordinates. This is admirable and a more stringent test of these
methods. We aim to investigate the issue of
initial search location further in subsequent work.

We conclude from this study that, for clusters with masses above
$10^{14}\,M_\odot$, the uncertainty in the methods seems to be around
a factor of two. Richness-based methods have the smallest
uncertainties, but this reliability may be due to the underlying simplicity of the HOD model, which includes no a-sphericity, dynamical substructure or large scale velocity distortions. However, we note that low scatter in the richness--mass relation has been observed for photometric samples (e.g., \citealt{2014arXiv1401.7716R}). Below $\rm 10^{14}\,M_\odot$, the scatter
rises as the number of member galaxies drops, and the uncertainty
rapidly approaches an order of magnitude. This level of error has
severe implications for studies of cosmology based on cluster masses
given the steeply-falling cluster mass function: there are many more
$\rm 10^{13}\,M_\odot$ clusters than $\rm 10^{14}\,M_\odot$ clusters
such that a large scatter in mass estimates will introduce very unpleasant
Malmquist-like biases that will render the answers meaningless unless the
biases can be very well modelled and controlled.

\indent In order to pinpoint the primary source of the errors, we also
supplied the participants with the ``true'' galaxy cluster membership,
as the halo has been initially populated by the HOD model. We then
asked the participants to return the group/cluster properties based on
this galaxy list rather than the one they had calculated. This
simplification did not improve mass estimates; for the majority of methods, the level of scatter was increased. The key factor here is the way in which methods have
been calibrated. Those which have been tuned to return unbiased results
on the basis of galaxies lying within the 3D `virial' radius will naturally
perform best when provided with such data, whilst methods attuned to
the more practical situation in which interlopers cannot be avoided
have adopted a variety of approaches to deal with this (aperture
selection, background subtraction etc.) and are likely to perform worse in the absence of the expected interlopers. We note that the masses of the cluster sample used for this `known' membership test are, on average, slightly lower than the `unknown' membership test. This may deliver a small contribution to the higher levels of scatter, as we have seen previously, that the level of scatter is higher for the lower mass clusters.

\indent The bottom line is that, with the
exception of the richness-based methods whose accuracy is unlikely to
be realised in a more realistic scenario, the limited number of
cluster tracers for the lower-mass systems (typically only $\sim 10-20$)
results in an irreducible large uncertainty in the cluster mass
estimate. We stress that this experiment has been carried out on the
most unchallenging possible test case of spherical systems with known locations and no imposed substructure. Observational 
challenges such as spectroscopic target selection, 
incompleteness, and slit/fibre collisions are also not considered. With a more realistic model for the galaxy population and a more observationally challenging set-up, it is likely that accurate group/cluster mass reconstruction will be even more problematic.
\section*{Acknowledgments}
The Millennium Simulation used in this paper was carried out by the
Virgo Supercomputing Consortium at the Computing Centre of the
Max-Planck Society in Garching. The halo merger trees used in the
paper are publicly available through the GAVO interface, found at
http://www.mpa-garching.mpg.de/millennium/. We would like to acknowledge funding from the Science and Technology Facilities Council (STFC). DC would like to thank the
Australian Research Council for receipt of a QEII Research
Fellowship. The Dark Cosmology Centre is funded by the Danish National Research Foundation. SIM acknowledges
the support of the STFC Studentship Enhancement Program (STEP). E.T. acknowledges the ESF grant MJD272. YOW acknowledges the support of the EU LaceGal grant: PIRSES-GA-2010-269264.\\
\indent The authors contributed in the following ways to this paper: LO, RAS,
FRP, \& DC designed and organised this project. LO performed the analysis presented and
wrote the majority of the paper. LO, ET,
SIM, MEG, RP, TP \& FRP organised the workshop that initiated this
project. MRM \& YW contributed to the analysis. The other authors (as
listed in section 3) provided results and descriptions of their
respective algorithms. All authors helped proof-read the paper.
\bibliographystyle{mn2e}
\bibliography{HMRC_phase1}

\begin{thebibliography}{109}
\expandafter\ifx\csname natexlab\endcsname\relax\def\natexlab#1{#1}\fi

\bibitem[{{Abell}(1958)}]{1958ApJS....3..211A}
{Abell} G.~O., 1958, \apjs, 3, 211

\bibitem[{{Albrecht} {et~al}\mbox{.}(2006){Albrecht}, {Bernstein}, {Cahn},
  {Freedman}, {Hewitt}, {Hu}, {Huth}, {Kamionkowski}, {Kolb}, {Knox}, {Mather},
  {Staggs}, \& {Suntzeff}}]{2006astro.ph..9591A}
{Albrecht} A. {et~al.}, 2006, ArXiv e-prints, astro-ph/0609591

\bibitem[{{Allen} {et~al}\mbox{.}(2011){Allen}, {Evrard}, \&
  {Mantz}}]{2011ARA&A..49..409A}
{Allen} S.~W., {Evrard} A.~E., {Mantz} A.~B., 2011, \araa, 49, 409

\bibitem[{{Applegate} {et~al}\mbox{.}(2012){Applegate}, {von der Linden},
  {Kelly}, {Allen}, {Allen}, {Burchat}, {Burke}, {Ebeling}, {Mantz}, \&
  {Morris}}]{2012arXiv1208.0605A}
{Applegate} D.~E. {et~al.}, 2012, ArXiv e-prints, 1208.0605

\bibitem[{{Ascaso} {et~al}\mbox{.}(2012){Ascaso}, {Wittman}, \&
  {Ben{\'{\i}}tez}}]{2012MNRAS.420.1167A}
{Ascaso} B., {Wittman} D., {Ben{\'{\i}}tez} N., 2012, \mnras, 420, 1167

\bibitem[{{Bahcall}(1988)}]{1988ARA&A..26..631B}
{Bahcall} N.~A., 1988, \araa, 26, 631

\bibitem[{{Bartelmann}(1996)}]{1996A&A...313..697B}
{Bartelmann} M., 1996, \aap, 313, 697

\bibitem[{{Beers} {et~al}\mbox{.}(1990){Beers}, {Flynn}, \&
  {Gebhardt}}]{1990AJ....100...32B}
{Beers} T.~C., {Flynn} K., {Gebhardt} K., 1990, \aj, 100, 32

\bibitem[{{Beraldo} {et~al}\mbox{.}(2013){Beraldo}, {Mamon}, {Duarte},
  {Peirani}, \& {Bou{\'e}}}]{2013arXiv1310.6756B}
{Beraldo} L.~J., {Mamon} G.~A., {Duarte} M., {Peirani} S., {Bou{\'e}} G., 2013,
  ArXiv e-prints, 1310.6756

\bibitem[{{Berlind} {et~al}\mbox{.}(2006){Berlind}, {Frieman}, {Weinberg},
  {Blanton}, {Warren}, {Abazajian}, {Scranton}, {Hogg}, {Scoccimarro},
  {Bahcall}, {Brinkmann}, {Gott}, {Kleinman}, {Krzesinski}, {Lee}, {Miller},
  {Nitta}, {Schneider}, {Tucker}, {Zehavi}, \& {SDSS
  Collaboration}}]{2006ApJS..167....1B}
{Berlind} A.~A. {et~al.}, 2006, \apjs, 167, 1

\bibitem[{{Biviano} {et~al}\mbox{.}(2006){Biviano}, {Murante}, {Borgani},
  {Diaferio}, {Dolag}, \& {Girardi}}]{2006A&A...456...23B}
{Biviano} A., {Murante} G., {Borgani} S., {Diaferio} A., {Dolag} K., {Girardi}
  M., 2006, \aap, 456, 23

\bibitem[{{Blanton} {et~al}\mbox{.}(2003){Blanton}, {Hogg}, \&
  {Bahcall}}]{2003ApJ...592..819B}
{Blanton} M.~R., {Hogg} D.~W., {Bahcall}, 2003, \apj, 592, 819

\bibitem[{{B{\"o}hringer} {et~al}\mbox{.}(2000){B{\"o}hringer}, {Voges},
  {Huchra}, {McLean}, {Giacconi}, {Rosati}, {Burg}, {Mader}, {Schuecker},
  {Simi{\c c}}, {Komossa}, {Reiprich}, {Retzlaff}, \&
  {Tr{\"u}mper}}]{2000ApJS..129..435B}
{B{\"o}hringer} H. {et~al.}, 2000, \apjs, 129, 435

\bibitem[{{Borgani} {et~al}\mbox{.}(1997){Borgani}, {Gardini}, {Girardi}, \&
  {Gottlober}}]{1997NewA....2..119B}
{Borgani} S., {Gardini} A., {Girardi} M., {Gottlober} S., 1997, \na, 2, 119

\bibitem[{{Carlstrom} {et~al}\mbox{.}(2002){Carlstrom}, {Holder}, \&
  {Reese}}]{2002ARA&A..40..643C}
{Carlstrom} J.~E., {Holder} G.~P., {Reese} E.~D., 2002, \araa, 40, 643

\bibitem[{{Cen}(1997)}]{1997ApJ...485...39C}
{Cen} R., 1997, \apj, 485, 39

\bibitem[{{Croton} {et~al}\mbox{.}(2006){Croton}, {Springel}, {White}, {De
  Lucia}, {Frenk}, {Gao}, {Jenkins}, {Kauffmann}, {Navarro}, \&
  {Yoshida}}]{2006MNRAS.365...11C}
{Croton} D.~J. {et~al.}, 2006, \mnras, 365, 11

\bibitem[{{Diaferio}(1999)}]{1999MNRAS.309..610D}
{Diaferio} A., 1999, \mnras, 309, 610

\bibitem[{{Diaferio} \& {Geller}(1997)}]{1997ApJ...481..633D}
{Diaferio} A., {Geller} M.~J., 1997, \apj, 481, 633

\bibitem[{{Duffy} {et~al}\mbox{.}(2008){Duffy}, {Schaye}, {Kay}, \& {Dalla
  Vecchia}}]{2008MNRAS.390L..64D}
{Duffy} A.~R., {Schaye} J., {Kay} S.~T., {Dalla Vecchia} C., 2008, \mnras, 390,
  L64

\bibitem[{{Einasto} {et~al}\mbox{.}(2001){Einasto}, {Einasto}, {Tago},
  {M{\"u}ller}, \& {Andernach}}]{2001AJ....122.2222E}
{Einasto} M., {Einasto} J., {Tago} E., {M{\"u}ller} V., {Andernach} H., 2001,
  \aj, 122, 2222

\bibitem[{{Eke} {et~al}\mbox{.}(2004){Eke}, {Baugh}, {Cole}, {Frenk},
  {Norberg}, {Peacock}, {Baldry}, {Bland-Hawthorn}, {Bridges}, {Cannon},
  {Colless}, {Collins}, {Couch}, {Dalton}, {de Propris}, {Driver},
  {Efstathiou}, {Ellis}, {Glazebrook}, {Jackson}, {Lahav}, {Lewis}, {Lumsden},
  {Maddox}, {Madgwick}, {Peterson}, {Sutherland}, \&
  {Taylor}}]{2004MNRAS.348..866E}
{Eke} V.~R. {et~al.}, 2004, \mnras, 348, 866

\bibitem[{Evrard {et~al}\mbox{.}(2008)Evrard, Bialek, Busha, White, Habib,
  Heitmann, Warren, Rasia, Tormen, \& Moscardini}]{Evrard:2008vo}
Evrard A.~E. {et~al.}, 2008, \apjs, 672, 122

\bibitem[{{Fadda} {et~al}\mbox{.}(1996){Fadda}, {Girardi}, {Giuricin},
  {Mardirossian}, \& {Mezzetti}}]{1996ApJ...473..670F}
{Fadda} D., {Girardi} M., {Giuricin} G., {Mardirossian} F., {Mezzetti} M.,
  1996, \apj, 473, 670

\bibitem[{{Forman} {et~al}\mbox{.}(1972){Forman}, {Kellogg}, {Gursky},
  {Tananbaum}, \& {Giacconi}}]{1972ApJ...178..309F}
{Forman} W., {Kellogg} E., {Gursky} H., {Tananbaum} H., {Giacconi} R., 1972,
  \apj, 178, 309

\bibitem[{{Gerke} {et~al}\mbox{.}(2013){Gerke}, {Wechsler}, {Behroozi},
  {Cooper}, {Yan}, \& {Coil}}]{2013ApJS..208....1G}
{Gerke} B.~F., {Wechsler} R.~H., {Behroozi} P.~S., {Cooper} M.~C., {Yan} R.,
  {Coil} A.~L., 2013, \apjs, 208, 1

\bibitem[{{Gifford} {et~al}\mbox{.}(2013){Gifford}, {Miller}, \&
  {Kern}}]{2013ApJ...773..116G}
{Gifford} D., {Miller} C., {Kern} N., 2013, \apj, 773, 116

\bibitem[{{Gifford} \& {Miller}(2013)}]{2013ApJ...768L..32G}
{Gifford} D., {Miller} C.~J., 2013, \apjl, 768, L32

\bibitem[{{Girardi} {et~al}\mbox{.}(1993){Girardi}, {Biviano}, {Giuricin},
  {Mardirossian}, \& {Mezzetti}}]{1993ApJ...404...38G}
{Girardi} M., {Biviano} A., {Giuricin} G., {Mardirossian} F., {Mezzetti} M.,
  1993, \apj, 404, 38

\bibitem[{{Gladders} \& {Yee}(2000)}]{2000AJ....120.2148G}
{Gladders} M.~D., {Yee} H.~K.~C., 2000, \aj, 120, 2148

\bibitem[{{Gladders} \& {Yee}(2005)}]{2005yCat..21570001G}
{Gladders} M.~D., {Yee} H.~K.~C., 2005, VizieR Online Data Catalog, 215, 70001

\bibitem[{{Goto} {et~al}\mbox{.}(2003){Goto}, {Yamauchi}, {Fujita}, {Okamura},
  {Sekiguchi}, {Smail}, {Bernardi}, \& {Gomez}}]{2003MNRAS.346..601G}
{Goto} T., {Yamauchi} C., {Fujita} Y., {Okamura} S., {Sekiguchi} M., {Smail}
  I., {Bernardi} M., {Gomez} P.~L., 2003, \mnras, 346, 601

\bibitem[{{Hasselfield} {et~al}\mbox{.}(2013){Hasselfield}, {Hilton},
  {Marriage}, {Addison}, {Barrientos}, {Battaglia}, {Battistelli}, {Bond},
  {Crichton}, {Das}, {Devlin}, {Dicker}, {Dunkley}, {D{\"u}nner}, {Fowler},
  {Gralla}, {Hajian}, {Halpern}, {Hincks}, {Hlozek}, {Hughes}, {Infante},
  {Irwin}, {Kosowsky}, {Marsden}, {Menanteau}, {Moodley}, {Niemack}, {Nolta},
  {Page}, {Partridge}, {Reese}, {Schmitt}, {Sehgal}, {Sherwin}, {Sievers},
  {Sif{\'o}n}, {Spergel}, {Staggs}, {Swetz}, {Switzer}, {Thornton}, {Trac}, \&
  {Wollack}}]{2013JCAP...07..008H}
{Hasselfield} M. {et~al.}, 2013, \jcap, 7, 8

\bibitem[{{Hearin} \& {Watson}(2013)}]{2013MNRAS.435.1313H}
{Hearin} A.~P., {Watson} D.~F., 2013, \mnras, 435, 1313

\bibitem[{{Huchra} \& {Geller}(1982)}]{1982ApJ...257..423H}
{Huchra} J.~P., {Geller} M.~J., 1982, \apj, 257, 423

\bibitem[{{Jian} {et~al}\mbox{.}(2013){Jian}, {Lin}, {Chiueh}, {Lin}, {Liu},
  {Merson}, {Baugh}, {Huang}, {Chen}, {Foucaud}, {Murphy}, {Cole}, {Burgett},
  \& {Kaiser}}]{2013arXiv1305.1891J}
{Jian} H.-Y. {et~al.}, 2013, ArXiv e-prints, 1305.1891

\bibitem[{{Kepner} {et~al}\mbox{.}(1999){Kepner}, {Fan}, {Bahcall}, {Gunn},
  {Lupton}, \& {Xu}}]{1999ApJ...517...78K}
{Kepner} J., {Fan} X., {Bahcall} N., {Gunn} J., {Lupton} R., {Xu} G., 1999,
  \apj, 517, 78

\bibitem[{{Knebe} {et~al}\mbox{.}(2011){Knebe}, {Knollmann}, {Muldrew},
  {Pearce}, {Aragon-Calvo}, {Ascasibar}, {Behroozi}, {Ceverino}, {Colombi},
  {Diemand}, {Dolag}, {Falck}, {Fasel}, {Gardner}, {Gottl{\"o}ber}, {Hsu},
  {Iannuzzi}, {Klypin}, {Luki{\'c}}, {Maciejewski}, {McBride}, {Neyrinck},
  {Planelles}, {Potter}, {Quilis}, {Rasera}, {Read}, {Ricker}, {Roy},
  {Springel}, {Stadel}, {Stinson}, {Sutter}, {Turchaninov}, {Tweed}, {Yepes},
  \& {Zemp}}]{2011MNRAS.415.2293K}
{Knebe} A. {et~al.}, 2011, \mnras, 415, 2293

\bibitem[{{Koester} {et~al}\mbox{.}(2007){Koester}, {McKay}, {Annis},
  {Wechsler}, {Evrard}, {Rozo}, {Bleem}, {Sheldon}, \&
  {Johnston}}]{2007ApJ...660..221K}
{Koester} B.~P. {et~al.}, 2007, \apj, 660, 221

\bibitem[{{Kravtsov} {et~al}\mbox{.}(2004){Kravtsov}, {Berlind}, {Wechsler},
  {Klypin}, {Gottl{\"o}ber}, {Allgood}, \& {Primack}}]{2004ApJ...609...35K}
{Kravtsov} A.~V., {Berlind} A.~A., {Wechsler} R.~H., {Klypin} A.~A.,
  {Gottl{\"o}ber} S., {Allgood} B., {Primack} J.~R., 2004, \apj, 609, 35

\bibitem[{{Li} \& {Yee}(2008)}]{2008AJ....135..809L}
{Li} I.~H., {Yee} H.~K.~C., 2008, \aj, 135, 809

\bibitem[{{Lopes} {et~al}\mbox{.}(2004){Lopes}, {de Carvalho}, {Gal},
  {Djorgovski}, {Odewahn}, {Mahabal}, \& {Brunner}}]{2004AJ....128.1017L}
{Lopes} P.~A.~A., {de Carvalho} R.~R., {Gal} R.~R., {Djorgovski} S.~G.,
  {Odewahn} S.~C., {Mahabal} A.~A., {Brunner} R.~J., 2004, \aj, 128, 1017

\bibitem[{{Lucey}(1983)}]{1983MNRAS.204...33L}
{Lucey} J.~R., 1983, \mnras, 204, 33

\bibitem[{{Macci{\`o}} {et~al}\mbox{.}(2008){Macci{\`o}}, {Dutton}, \& {van den
  Bosch}}]{2008MNRAS.391.1940M}
{Macci{\`o}} A.~V., {Dutton} A.~A., {van den Bosch} F.~C., 2008, \mnras, 391,
  1940

\bibitem[{{Mamon} {et~al}\mbox{.}(2013){Mamon}, {Biviano}, \&
  {Bou{\'e}}}]{2013MNRAS.429.3079M}
{Mamon} G.~A., {Biviano} A., {Bou{\'e}} G., 2013, \mnras, 429, 3079

\bibitem[{{Mamon} {et~al}\mbox{.}(2010){Mamon}, {Biviano}, \&
  {Murante}}]{2010A&A...520A..30M}
{Mamon} G.~A., {Biviano} A., {Murante} G., 2010, \aap, 520, A30

\bibitem[{{Mamon} \& {{\L}okas}(2005)}]{2005MNRAS.363..705M}
{Mamon} G.~A., {{\L}okas} E.~L., 2005, \mnras, 363, 705

\bibitem[{{Marinoni} {et~al}\mbox{.}(2002){Marinoni}, {Davis}, {Newman}, \&
  {Coil}}]{2002ApJ...580..122M}
{Marinoni} C., {Davis} M., {Newman} J.~A., {Coil} A.~L., 2002, \apj, 580, 122

\bibitem[{{Mart{\'{\i}}nez} {et~al}\mbox{.}(2008){Mart{\'{\i}}nez}, {Coenda},
  \& {Muriel}}]{2008MNRAS.391..585M}
{Mart{\'{\i}}nez} H.~J., {Coenda} V., {Muriel} H., 2008, \mnras, 391, 585

\bibitem[{{Masaki} {et~al}\mbox{.}(2013){Masaki}, {Lin}, \&
  {Yoshida}}]{2013MNRAS.436.2286M}
{Masaki} S., {Lin} Y.-T., {Yoshida} N., 2013, \mnras, 436, 2286

\bibitem[{{Menanteau} {et~al}\mbox{.}(2009){Menanteau}, {Hughes}, {Jimenez},
  {Hernandez-Monteagudo}, {Verde}, {Kosowsky}, {Moodley}, {Infante}, \&
  {Roche}}]{2009ApJ...698.1221M}
{Menanteau} F. {et~al.}, 2009, \apj, 698, 1221

\bibitem[{{Milvang-Jensen} {et~al}\mbox{.}(2008){Milvang-Jensen}, {Noll},
  {Halliday}, {Poggianti}, {Jablonka}, {Arag{\'o}n-Salamanca}, {Saglia},
  {Nowak}, {von der Linden}, {De Lucia}, {Pell{\'o}}, {Moustakas}, {Poirier},
  {Bamford}, {Clowe}, {Dalcanton}, {Rudnick}, {Simard}, {White}, \&
  {Zaritsky}}]{2008A&A...482..419M}
{Milvang-Jensen} B. {et~al.}, 2008, \aap, 482, 419

\bibitem[{{Mu{\~n}oz-Cuartas} \& {M{\"u}ller}(2012)}]{2012MNRAS.423.1583M}
{Mu{\~n}oz-Cuartas} J.~C., {M{\"u}ller} V., 2012, \mnras, 423, 1583

\bibitem[{{Muldrew} {et~al}\mbox{.}(2012){Muldrew}, {Croton}, {Skibba},
  {Pearce}, {Ann}, {Baldry}, {Brough}, {Choi}, {Conselice}, {Cowan},
  {Gallazzi}, {Gray}, {Gr{\"u}tzbauch}, {Li}, {Park}, {Pilipenko}, {Podgorzec},
  {Robotham}, {Wilman}, {Yang}, {Zhang}, \& {Zibetti}}]{2012MNRAS.419.2670M}
{Muldrew} S.~I. {et~al.}, 2012, \mnras, 419, 2670

\bibitem[{{Muldrew} {et~al}\mbox{.}(2011){Muldrew}, {Pearce}, \&
  {Power}}]{2011MNRAS.410.2617M}
{Muldrew} S.~I., {Pearce} F.~R., {Power} C., 2011, \mnras, 410, 2617

\bibitem[{{Munari} {et~al}\mbox{.}(2013){Munari}, {Biviano}, {Borgani},
  {Murante}, \& {Fabjan}}]{2013MNRAS.430.2638M}
{Munari} E., {Biviano} A., {Borgani} S., {Murante} G., {Fabjan} D., 2013,
  \mnras, 430, 2638

\bibitem[{{Murphy} {et~al}\mbox{.}(2012){Murphy}, {Geach}, \&
  {Bower}}]{2012MNRAS.420.1861M}
{Murphy} D.~N.~A., {Geach} J.~E., {Bower} R.~G., 2012, \mnras, 420, 1861

\bibitem[{{Navarro} {et~al}\mbox{.}(1996){Navarro}, {Frenk}, \&
  {White}}]{1996ApJ...462..563N}
{Navarro} J.~F., {Frenk} C.~S., {White} S.~D.~M., 1996, \apj, 462, 563

\bibitem[{{Navarro} {et~al}\mbox{.}(1997){Navarro}, {Frenk}, \&
  {White}}]{1997ApJ...490..493N}
{Navarro} J.~F., {Frenk} C.~S., {White} S.~D.~M., 1997, \apj, 490, 493

\bibitem[{{Old} {et~al}\mbox{.}(2013){Old}, {Gray}, \&
  {Pearce}}]{2013MNRAS.434.2606O}
{Old} L., {Gray} M.~E., {Pearce} F.~R., 2013, \mnras, 434, 2606

\bibitem[{{Olsen} {et~al}\mbox{.}(1999){Olsen}, {Scodeggio}, {da Costa},
  {Benoist}, {Bertin}, {Deul}, {Erben}, {Guarnieri}, {Hook}, {Nonino},
  {Prandoni}, {Slijkhuis}, {Wicenec}, \& {Wichmann}}]{1999A&A...345..681O}
{Olsen} L.~F. {et~al.}, 1999, \aap, 345, 681

\bibitem[{{Papovich}(2008)}]{2008ApJ...676..206P}
{Papovich} C., 2008, \apj, 676, 206

\bibitem[{{Phleps} {et~al}\mbox{.}(2013){Phleps}, {Wilman}, {Zibetti}, \&
  {Budav{\'a}ri}}]{2013arXiv1312.1340P}
{Phleps} S., {Wilman} D.~J., {Zibetti} S., {Budav{\'a}ri} T., 2013, ArXiv
  e-prints, 1313.1340

\bibitem[{{Planck Collaboration} {et~al}\mbox{.}(2013){Planck Collaboration},
  {Ade}, {Aghanim}, {Armitage-Caplan}, {Arnaud}, {Ashdown}, {Atrio-Barandela},
  {Aumont}, {Aussel}, {Baccigalupi}, \& et~al.}]{2013arXiv1303.5089P}
{Planck Collaboration} {et~al.}, 2013, ArXiv e-prints, 1303.5089

\bibitem[{{Postman} {et~al}\mbox{.}(2005){Postman}, {Franx}, {Cross}, {Holden},
  {Ford}, {Illingworth}, {Goto}, {Demarco}, {Rosati}, {Blakeslee}, {Tran},
  {Ben{\'{\i}}tez}, {Clampin}, {Hartig}, {Homeier}, {Ardila}, {Bartko},
  {Bouwens}, {Bradley}, {Broadhurst}, {Brown}, {Burrows}, {Cheng}, {Feldman},
  {Golimowski}, {Gronwall}, {Infante}, {Kimble}, {Krist}, {Lesser}, {Martel},
  {Mei}, {Menanteau}, {Meurer}, {Miley}, {Motta}, {Sirianni}, {Sparks}, {Tran},
  {Tsvetanov}, {White}, \& {Zheng}}]{2005ApJ...623..721P}
{Postman} M. {et~al.}, 2005, \apj, 623, 721

\bibitem[{{Postman} {et~al}\mbox{.}(1996){Postman}, {Lubin}, {Gunn}, {Oke},
  {Hoessel}, {Schneider}, \& {Christensen}}]{1996AJ....111..615P}
{Postman} M., {Lubin} L.~M., {Gunn} J.~E., {Oke} J.~B., {Hoessel} J.~G.,
  {Schneider} D.~P., {Christensen} J.~A., 1996, \aj, 111, 615

\bibitem[{{Rosati} {et~al}\mbox{.}(2002){Rosati}, {Borgani}, \&
  {Norman}}]{2002ARA&A..40..539R}
{Rosati} P., {Borgani} S., {Norman} C., 2002, \araa, 40, 539

\bibitem[{{Rozo} {et~al}\mbox{.}(2014){Rozo}, {Rykoff}, {Bartlett}, \&
  {Melin}}]{2014arXiv1401.7716R}
{Rozo} E., {Rykoff} E.~S., {Bartlett} J.~G., {Melin} J.~B., 2014, ArXiv
  e-prints, 1401.7716

\bibitem[{{Rykoff} {et~al}\mbox{.}(2013){Rykoff}, {Rozo}, {Busha}, {Cunha},
  {Finoguenov}, {Evrard}, {Hao}, {Koester}, {Leauthaud}, {Nord}, {Pierre},
  {Reddick}, {Sadibekova}, {Sheldon}, \& {Wechsler}}]{2013arXiv1303.3562R}
{Rykoff} E.~S. {et~al.}, 2013, ArXiv e-prints, 1303.3562

\bibitem[{{Sanderson} \& {Ponman}(2010)}]{2010MNRAS.402...65S}
{Sanderson} A.~J.~R., {Ponman} T.~J., 2010, \mnras, 402, 65

\bibitem[{{Saro} {et~al}\mbox{.}(2013){Saro}, {Mohr}, {Bazin}, \&
  {Dolag}}]{2013ApJ...772...47S}
{Saro} A., {Mohr} J.~J., {Bazin} G., {Dolag} K., 2013, \apj, 772, 47

\bibitem[{{Serra} {et~al}\mbox{.}(2011){Serra}, {Diaferio}, {Murante}, \&
  {Borgani}}]{2011MNRAS.412..800S}
{Serra} A.~L., {Diaferio} A., {Murante} G., {Borgani} S., 2011, \mnras, 412,
  800

\bibitem[{{Sheth} \& {Diaferio}(2001)}]{2001MNRAS.322..901S}
{Sheth} R.~K., {Diaferio} A., 2001, \mnras, 322, 901

\bibitem[{Sifon {et~al}\mbox{.}(2013)Sifon, Menanteau, Hasselfield, Marriage,
  Hughes, Barrientos, Gonzalez, Infante, Addison, Baker, Battaglia, Bond,
  Crichton, Das, Devlin, Dunkley, Dunner, Gralla, Hajian, Hilton, Hincks,
  Kosowsky, Marsden, Moodley, Niemack, Nolta, Page, Partridge, Reese, Sehgal,
  Sievers, Spergel, Staggs, Thornton, Trac, \& Wollack}]{2013ApJ...772...25S}
Sifon C. {et~al.}, 2013, \apj, 772, 25

\bibitem[{{Skibba} {et~al}\mbox{.}(2006){Skibba}, {Sheth}, {Connolly}, \&
  {Scranton}}]{2006MNRAS.369...68S}
{Skibba} R., {Sheth} R.~K., {Connolly} A.~J., {Scranton} R., 2006, \mnras, 369,
  68

\bibitem[{{Skibba}(2009)}]{2009MNRAS.392.1467S}
{Skibba} R.~A., 2009, \mnras, 392, 1467

\bibitem[{{Skibba} \& {Sheth}(2009)}]{2009MNRAS.392.1080S}
{Skibba} R.~A., {Sheth} R.~K., 2009, \mnras, 392, 1080

\bibitem[{{Skibba} {et~al}\mbox{.}(2011){Skibba}, {van den Bosch}, {Yang},
  {More}, {Mo}, \& {Fontanot}}]{2011MNRAS.410..417S}
{Skibba} R.~A., {van den Bosch} F.~C., {Yang} X., {More} S., {Mo} H.,
  {Fontanot} F., 2011, \mnras, 410, 417

\bibitem[{{Soares-Santos} {et~al}\mbox{.}(2011){Soares-Santos}, {de Carvalho},
  {Annis}, {Gal}, {La Barbera}, {Lopes}, {Wechsler}, {Busha}, \&
  {Gerke}}]{2011ApJ...727...45S}
{Soares-Santos} M. {et~al.}, 2011, \apj, 727, 45

\bibitem[{{Song} {et~al}\mbox{.}(2012){Song}, {Mohr}, {Barkhouse}, {Warren},
  {Dolag}, \& {Rude}}]{2012ApJ...747...58S}
{Song} J., {Mohr} J.~J., {Barkhouse} W.~A., {Warren} M.~S., {Dolag} K., {Rude}
  C., 2012, \apj, 747, 58

\bibitem[{{Springel} {et~al}\mbox{.}(2005){Springel}, {White}, {Jenkins},
  {Frenk}, {Yoshida}, {Gao}, {Navarro}, {Thacker}, {Croton}, {Helly},
  {Peacock}, {Cole}, {Thomas}, {Couchman}, {Evrard}, {Colberg}, \&
  {Pearce}}]{2005Natur.435..629S}
{Springel} V. {et~al.}, 2005, \nat, 435, 629

\bibitem[{{Springel} {et~al}\mbox{.}(2001){Springel}, {White}, {Tormen}, \&
  {Kauffmann}}]{2001MNRAS.328..726S}
{Springel} V., {White} S.~D.~M., {Tormen} G., {Kauffmann} G., 2001, \mnras,
  328, 726

\bibitem[{{Sun} {et~al}\mbox{.}(2009){Sun}, {Voit}, {Donahue}, {Jones},
  {Forman}, \& {Vikhlinin}}]{2009ApJ...693.1142S}
{Sun} M., {Voit} G.~M., {Donahue} M., {Jones} C., {Forman} W., {Vikhlinin} A.,
  2009, \apj, 693, 1142

\bibitem[{{Sunyaev} \& {Zeldovich}(1972)}]{1972CoASP...4..173S}
{Sunyaev} R.~A., {Zeldovich} Y.~B., 1972, \coasp, 4, 173

\bibitem[{{Szabo} {et~al}\mbox{.}(2011){Szabo}, {Pierpaoli}, {Dong}, {Pipino},
  \& {Gunn}}]{2011ApJ...736...21S}
{Szabo} T., {Pierpaoli} E., {Dong} F., {Pipino} A., {Gunn} J., 2011, \apj, 736,
  21

\bibitem[{{Tago} {et~al}\mbox{.}(2008){Tago}, {Einasto}, {Saar}, {Tempel},
  {Einasto}, {Vennik}, \& {M{\"u}ller}}]{2008A&A...479..927T}
{Tago} E., {Einasto} J., {Saar} E., {Tempel} E., {Einasto} M., {Vennik} J.,
  {M{\"u}ller} V., 2008, \aap, 479, 927

\bibitem[{{Tago} {et~al}\mbox{.}(2010){Tago}, {Saar}, {Tempel}, {Einasto},
  {Einasto}, {Nurmi}, \& {Hein{\"a}m{\"a}ki}}]{2010A&A...514A.102T}
{Tago} E., {Saar} E., {Tempel} E., {Einasto} J., {Einasto} M., {Nurmi} P.,
  {Hein{\"a}m{\"a}ki} P., 2010, \aap, 514, A102

\bibitem[{{Tempel} {et~al}\mbox{.}(2012){Tempel}, {Tago}, \&
  {Liivam{\"a}gi}}]{2012A&A...540A.106T}
{Tempel} E., {Tago} E., {Liivam{\"a}gi} L.~J., 2012, \aap, 540, A106

\bibitem[{{Tempel} {et~al}\mbox{.}(2014){Tempel}, {Tamm}, {Gramann},
  {Tuvikene}, {Liivam{\"a}gi}, {Suhhonenko}, {Kipper}, {Einasto}, \&
  {Saar}}]{2014arXiv1402.1350T}
{Tempel} E. {et~al.}, 2014, ArXiv e-prints, 1402.1350

\bibitem[{{van Breukelen} \& {Clewley}(2009)}]{2009MNRAS.395.1845V}
{van Breukelen} C., {Clewley} L., 2009, \mnras, 395, 1845

\bibitem[{{van den Bosch} {et~al}\mbox{.}(2008){van den Bosch}, {Aquino},
  {Yang}, {Mo}, {Pasquali}, {McIntosh}, {Weinmann}, \&
  {Kang}}]{2008MNRAS.387...79V}
{van den Bosch} F.~C., {Aquino} D., {Yang} X., {Mo} H.~J., {Pasquali} A.,
  {McIntosh} D.~H., {Weinmann} S.~M., {Kang} X., 2008, \mnras, 387, 79

\bibitem[{{Vanderlinde} {et~al}\mbox{.}(2010){Vanderlinde}, {Crawford}, {de
  Haan}, {Dudley}, {Shaw}, {Ade}, {Aird}, {Benson}, {Bleem}, {Brodwin},
  {Carlstrom}, {Chang}, {Crites}, {Desai}, {Dobbs}, {Foley}, {George},
  {Gladders}, {Hall}, {Halverson}, {High}, {Holder}, {Holzapfel}, {Hrubes},
  {Joy}, {Keisler}, {Knox}, {Lee}, {Leitch}, {Loehr}, {Lueker}, {Marrone},
  {McMahon}, {Mehl}, {Meyer}, {Mohr}, {Montroy}, {Ngeow}, {Padin}, {Plagge},
  {Pryke}, {Reichardt}, {Rest}, {Ruel}, {Ruhl}, {Schaffer}, {Shirokoff},
  {Song}, {Spieler}, {Stalder}, {Staniszewski}, {Stark}, {Stubbs}, {van
  Engelen}, {Vieira}, {Williamson}, {Yang}, {Zahn}, \&
  {Zenteno}}]{2010ApJ...722.1180V}
{Vanderlinde} K. {et~al.}, 2010, \apj, 722, 1180

\bibitem[{{Vikhlinin} {et~al}\mbox{.}(2009){Vikhlinin}, {Burenin}, {Ebeling},
  {Forman}, {Hornstrup}, {Jones}, {Kravtsov}, {Murray}, {Nagai}, {Quintana}, \&
  {Voevodkin}}]{2009ApJ...692.1033V}
{Vikhlinin} A. {et~al.}, 2009, \apj, 692, 1033

\bibitem[{{von der Linden} {et~al}\mbox{.}(2007){von der Linden}, {Best},
  {Kauffmann}, \& {White}}]{2007MNRAS.379..867V}
{von der Linden} A., {Best} P.~N., {Kauffmann} G., {White} S.~D.~M., 2007,
  \mnras, 379, 867

\bibitem[{{Warren} {et~al}\mbox{.}(2006){Warren}, {Abazajian}, {Holz}, \&
  {Teodoro}}]{2006ApJ...646..881W}
{Warren} M.~S., {Abazajian} K., {Holz} D.~E., {Teodoro} L., 2006, \apj, 646,
  881

\bibitem[{{Willis} {et~al}\mbox{.}(2013){Willis}, {Clerc}, {Bremer}, {Pierre},
  {Adami}, {Ilbert}, {Maughan}, {Maurogordato}, {Pacaud}, {Valtchanov},
  {Chiappetti}, {Thanjavur}, {Gwyn}, {Stanway}, \&
  {Winkworth}}]{2013MNRAS.430..134W}
{Willis} J.~P. {et~al.}, 2013, \mnras, 430, 134

\bibitem[{{Wojtak} {et~al}\mbox{.}(2009){Wojtak}, {{\L}okas}, {Mamon}, \&
  {Gottl{\"o}ber}}]{2009MNRAS.399..812W}
{Wojtak} R., {{\L}okas} E.~L., {Mamon} G.~A., {Gottl{\"o}ber} S., 2009, \mnras,
  399, 812

\bibitem[{{Wojtak} {et~al}\mbox{.}(2008){Wojtak}, {{\L}okas}, {Mamon},
  {Gottl{\"o}ber}, {Klypin}, \& {Hoffman}}]{2008MNRAS.388..815W}
{Wojtak} R., {{\L}okas} E.~L., {Mamon} G.~A., {Gottl{\"o}ber} S., {Klypin} A.,
  {Hoffman} Y., 2008, \mnras, 388, 815

\bibitem[{{Wojtak} {et~al}\mbox{.}(2007){Wojtak}, {{\L}okas}, {Mamon},
  {Gottl{\"o}ber}, {Prada}, \& {Moles}}]{2007A&A...466..437W}
{Wojtak} R., {{\L}okas} E.~L., {Mamon} G.~A., {Gottl{\"o}ber} S., {Prada} F.,
  {Moles} M., 2007, \aap, 466, 437

\bibitem[{{Yahil} \& {Vidal}(1977)}]{1977ApJ...214..347Y}
{Yahil} A., {Vidal} N.~V., 1977, \apj, 214, 347

\bibitem[{{Yang} {et~al}\mbox{.}(2005{\natexlab{a}}){Yang}, {Mo}, {van den
  Bosch}, \& {Jing}}]{2005MNRAS.356.1293Y}
{Yang} X., {Mo} H.~J., {van den Bosch} F.~C., {Jing} Y.~P., 2005{\natexlab{a}},
  \mnras, 356, 1293

\bibitem[{{Yang} {et~al}\mbox{.}(2005{\natexlab{b}}){Yang}, {Mo}, {van den
  Bosch}, \& {Jing}}]{2005MNRAS.357..608Y}
{Yang} X., {Mo} H.~J., {van den Bosch} F.~C., {Jing} Y.~P., 2005{\natexlab{b}},
  \mnras, 357, 608

\bibitem[{{Yee} \& {Ellingson}(2003)}]{2003ApJ...585..215Y}
{Yee} H.~K.~C., {Ellingson} E., 2003, \apj, 585, 215

\bibitem[{{York} {et~al}\mbox{.}(2000){York}, {Adelman}, {Anderson},
  {Anderson}, {Annis}, {Bahcall}, {Bakken}, {Barkhouser}, {Bastian}, {Berman},
  {Boroski}, {Bracker}, {Briegel}, {Briggs}, {Brinkmann}, {Brunner}, {Burles},
  {Carey}, {Carr}, {Castander}, {Chen}, {Colestock}, {Connolly}, {Crocker},
  {Csabai}, {Czarapata}, {Davis}, {Doi}, {Dombeck}, {Eisenstein}, {Ellman},
  {Elms}, {Evans}, {Fan}, {Federwitz}, {Fiscelli}, {Friedman}, {Frieman},
  {Fukugita}, {Gillespie}, {Gunn}, {Gurbani}, {de Haas}, {Haldeman}, {Harris},
  {Hayes}, {Heckman}, {Hennessy}, {Hindsley}, {Holm}, {Holmgren}, {Huang},
  {Hull}, {Husby}, {Ichikawa}, {Ichikawa}, {Ivezi{\'c}}, {Kent}, {Kim},
  {Kinney}, {Klaene}, {Kleinman}, {Kleinman}, {Knapp}, {Korienek}, {Kron},
  {Kunszt}, {Lamb}, {Lee}, {Leger}, {Limmongkol}, {Lindenmeyer}, {Long},
  {Loomis}, {Loveday}, {Lucinio}, {Lupton}, {MacKinnon}, {Mannery}, {Mantsch},
  {Margon}, {McGehee}, {McKay}, {Meiksin}, {Merelli}, {Monet}, {Munn},
  {Narayanan}, {Nash}, {Neilsen}, {Neswold}, {Newberg}, {Nichol}, {Nicinski},
  {Nonino}, {Okada}, {Okamura}, {Ostriker}, {Owen}, {Pauls}, {Peoples},
  {Peterson}, {Petravick}, {Pier}, {Pope}, {Pordes}, {Prosapio},
  {Rechenmacher}, {Quinn}, {Richards}, {Richmond}, {Rivetta}, {Rockosi},
  {Ruthmansdorfer}, {Sandford}, {Schlegel}, {Schneider}, {Sekiguchi}, {Sergey},
  {Shimasaku}, {Siegmund}, {Smee}, {Smith}, {Snedden}, {Stone}, {Stoughton},
  {Strauss}, {Stubbs}, {SubbaRao}, {Szalay}, {Szapudi}, {Szokoly}, {Thakar},
  {Tremonti}, {Tucker}, {Uomoto}, {Vanden Berk}, {Vogeley}, {Waddell}, {Wang},
  {Watanabe}, {Weinberg}, {Yanny}, {Yasuda}, \& {SDSS
  Collaboration}}]{2000AJ....120.1579Y}
{York} D.~G. {et~al.}, 2000, \aj, 120, 1579

\bibitem[{{Zehavi} {et~al}\mbox{.}(2011){Zehavi}, {Zheng}, {Weinberg},
  {Blanton}, {Bahcall}, {Berlind}, {Brinkmann}, {Frieman}, {Gunn}, {Lupton},
  {Nichol}, {Percival}, {Schneider}, {Skibba}, {Strauss}, {Tegmark}, \&
  {York}}]{2011ApJ...736...59Z}
{Zehavi} I. {et~al.}, 2011, \apj, 736, 59

\bibitem[{{Zehavi} {et~al}\mbox{.}(2005){Zehavi}, {Zheng}, {Weinberg},
  {Frieman}, {Berlind}, {Blanton}, {Scoccimarro}, {Sheth}, {Strauss}, {Kayo},
  {Suto}, {Fukugita}, {Nakamura}, {Bahcall}, {Brinkmann}, {Gunn}, {Hennessy},
  {Ivezi{\'c}}, {Knapp}, {Loveday}, {Meiksin}, {Schlegel}, {Schneider},
  {Szapudi}, {Tegmark}, {Vogeley}, {York}, \& {SDSS
  Collaboration}}]{2005ApJ...630....1Z}
{Zehavi} I. {et~al.}, 2005, \apj, 630, 1

\bibitem[{{Zheng} {et~al}\mbox{.}(2007){Zheng}, {Coil}, \&
  {Zehavi}}]{2007ApJ...667..760Z}
{Zheng} Z., {Coil} A.~L., {Zehavi} I., 2007, \apj, 667, 760

\bibitem[{{Zwicky}(1937)}]{1937ApJ....86..217Z}
{Zwicky} F., 1937, \apj, 86, 217

\bibitem[{{Zwicky} {et~al}\mbox{.}(1968){Zwicky}, {Herzog}, \&
  {Wild}}]{1968cgcg.book.....Z}
{Zwicky} F., {Herzog} E., {Wild} P., 1968, {Catalogue of galaxies and of
  clusters of galaxies}

\end{thebibliography}
\newpage
\begin{table*}

\appendix
\begin{flushleft}
\section{Properties of the Mass Reconstruction Methods}
\end{flushleft}
 \centering
 \caption{Table illustrating the member galaxy selection process for all methods. The second column details how each method selects an initial member galaxy sample, while the third column outlines the member galaxy sample refining process. Finally, the fourth column describes how methods treat interloping galaxies that are not associated with the clusters.}
 \begin{tabular}{c p{4.0cm} p{4.8cm} p{3.8cm}}
 \toprule
 \multirow{2}[3]{*}{\textbf{Methods}}&\multicolumn{3}{c}{Member galaxy selection methodology} \\[1.0ex]
 \cline{2-4}
 &Initial Galaxy Selection&Refine Membership&Treatment of Interlopers \\
 \midrule
 \textcolor{magenta}{\textbf{PCN}}&Within $\rm 5\,Mpc$, $\rm 1000\,km\,s^{-1}$&Clipping of $\pm3\,\sigma$, using galaxies within $\rm 1\,Mpc$&Use galaxies at $\rm 3-5 \,Mpc$ to find interloper population to remove \\
 \textcolor{magenta}{\textbf{PFN}}&FOF&No&No \\
 
 \textcolor{magenta}{\textbf{NUM}}&Within $\rm 1\,Mpc$, $\rm 1333\,km\,s^{-1}$&1) Estimate $R_{\rm 200}$ by a relationship between $R_{\rm 200}$ and richness deduced from CLE; 2) Galaxies within $R_{\rm 200}$ and with velocities less than $2.7\,\sigma_{\rm los}(R)$ are selected&No \\
\hline
 \textcolor{black}{\textbf{ESC}}&Within preliminary $R_{\rm 200}$ estimate and $\rm \pm3500 \,km\,s^{-1}$&Gapper technique&Removed in refining by Gapper technique \\
 
 \textcolor{black}{\textbf{MPO}}&Input from CLN&1) Calculate $R_{\rm 200}$, $R_{\rm \rho}$, $R_{\rm red}$, $R_{\rm blue}$ by MAMPOSSt method; 2) Select members within radius according to colour&No \\
 
 \textcolor{black}{\textbf{MP1}}&Input from CLN&Same as MPO except colour blind&No \\
 
 \textcolor{black}{\textbf{RW}}&Within $\rm 3\,\,Mpc$, $\rm 4000\,\,km\,s^{-1}$&Within $R_{\rm 200}$, $|2\Phi(R)|^{1/2}$, where $R_{\rm 200}$ obtained iteratively
&No \\

 \textcolor{black}{\textbf{TAR}}&FOF&No&No \\
 \hline
 \textcolor{blue}{\textbf{PCO}}&Input from PCN& Input from PCN&Include interloper contamination in density fitting \\
 
 \textcolor{blue}{\textbf{PFO}}&Input from PFN&Input from PFN&No \\
 
 \textcolor{blue}{\textbf{PCR}}&Input from PCN&Input from PCN&Same as PCN \\
 
 \textcolor{blue}{\textbf{PFR}}&Input from PFN&Input from PFN&No \\
 \hline
 \textcolor{green}{\textbf{HBM}}&FOF (ellipsoidal search range, centre of most luminous galaxy)&Increasing mass limits, then FOF, loops until closure condition&No \\
 
 \textcolor{green}{\textbf{MVM}}&Same as HBM&Same as HBM&No \\
 \hline
 \textcolor{red}{\textbf{AS1}}&Within $\rm 1\,Mpc$, $\rm 4000\,km\,s^{-1}$, constrained by colour-magnitude relation&Clipping of $\pm3\,\sigma$&Removed by clipping of $\pm3\,\sigma$ \\
 
 \textcolor{red}{\textbf{AS2}}&Within $\rm 1\,Mpc$, $4\rm 000\,km\,s^{-1}$, constrained by colour-magnitude relation&Clipping of $\pm3\,\sigma$&Removed by clipping of $\pm3\,\sigma$ \\
 
 \textcolor{red}{\textbf{AvL}}&Within $2.5\,\sigma_{v}$ and $0.8\,R_{\rm 200}$&Obtain $R_{\rm 200}$ and $\,\sigma_{v}$ by $\,\sigma$-clipping&No \\
 
 \textcolor{red}{\textbf{CLE}}&Within $\rm 3\,Mpc$ ,$\rm 4000\,km\,s^{-1}$&1) Estimate $R_{\rm 200}$ by aperture velocity dispersion; 2) galaxies within $R_{\rm 200}$ and with velocities less than 2.7$\,\sigma_{\rm los}(R)$ are selected; 3) Iterate steps 1 and 2 until convergence&Obvious interlopers are removed by velocity gap technique, then further treated in iteration by $\sigma$ clipping \\
 
 \textcolor{red}{\textbf{CLN}}&Input from NUM&Same as CLE&Same as CLE \\
 
 \textcolor{red}{\textbf{SG1}}&Within $\rm 4000 \,km\,s^{-1}$&1) Measure $\,\sigma_{\rm gal}$, estimate $M_{\rm 200}$ and $R_{\rm 200}$; 2) Select galaxies within $R_{\rm 200}$; 3) Iterate steps 1 and 2 until convergence&Shifting gapper with minimum bin size of $\rm 250 \,kpc$ and 15 galaxies; velocity limit $\rm 1000\,km\,s^{-1}$ from main body \\
 
 \textcolor{red}{\textbf{SG2}}&Within $\rm 4000 \,km\,s^{-1}$&1) Measure $\sigma_{\rm gal}$, estimate $M_{\rm 200}$ and $R_{\rm 200}$; 2) Select galaxies within $R_{\rm 200}$; 3) Iterate steps 1 and 2 until convergence&Shifting gapper with minimum bin size of $\rm 150 \,kpc$ and 10 galaxies; velocity limit $\rm 500 \,km\,s^{-1}$ from main body \\
 \textcolor{red}{\textbf{PCS}}&Input from PCN&Input from PCN&Same as PCN \\
 \textcolor{red}{\textbf{PFS}}&Input from PFN&Input from PFN&No \\
 \bottomrule
 \end{tabular}
 \label{table:appendix_table_1}
\end{table*}
\begin{table*}
\renewcommand\thetable{A2} 
 \centering
 \caption{Table showing the characteristics of the mass reconstruction process of methods used in this comparison. The second, third, fourth and fifth columns illustrate whether a method calculates/utilizes the velocities, velocity dispersion, radial distance of galaxies from cluster centre, the richness and the projected phase space information of galaxies respectively. If a method assumed a mass or number density profile it is indicated in columns six and seven.}
 \begin{tabular}{c p{1.0cm} p{1.3cm} p{1.2cm} p{1.0cm} p{1.4cm} p{1.9cm} p{1.7cm} p{1.5cm}}
 \toprule
 \multirow{2}[4]{*}{\textbf{Methods}}&\multicolumn{7}{c}{Properties used to recovering halo mass} \\[1.0ex]
 \cline{2-8}
 &Velocities&Velocity dispersion&Radial distance&Richness&Projected phase space&Mass density profile&Number density profile\\
 \hline
 \textcolor{magenta}{\textbf{PCN}}&Yes&No&No&Yes&No&No&No \\
 \textcolor{magenta}{\textbf{PFN}}&Yes&No&No&Yes&No&No&No\\
 \textcolor{magenta}{\textbf{NUM}}&Yes&No&No&Yes&Yes&No&No\\
  \hline
 \textcolor{black}{\textbf{ESC}}&Yes&Yes&Yes&No&No&Caustics&No \\
 \textcolor{black}{\textbf{MPO}}&Yes&No&Yes&No&Yes&NFW&Yes\\
 \textcolor{black}{\textbf{MP1}}&Yes&No&Yes&No&Yes&NFW&Yes\\
 \textcolor{black}{\textbf{RW}}&Yes&No&Yes&No&Yes&NFW&Yes\\
 \textcolor{black}{\textbf{TAR}}&Yes&Yes&Yes&No&No& NFW&No\\
 \hline
 \textcolor{blue}{\textbf{PCO}}&Yes&No&No&No&No&NFW&Yes\\
 \textcolor{blue}{\textbf{PFO}}&Yes&No&No&No&No&NFW&Yes\\
 \textcolor{blue}{\textbf{PCR}}&Yes&No&Yes&No&No&No&No\\
 \textcolor{blue}{\textbf{PFR}}&Yes&No&Yes&No&No&No&No \\
  \hline
\textcolor{green}{\textbf{HBM}}&Yes&Yes&Yes&No&No&NFW&No\\
 \textcolor{green}{\textbf{MVM}}&Yes&Yes&Yes&No&No&NFW&No\\
 \hline
 \textcolor{red}{\textbf{AS1}}&Yes&Yes&No&No&No&No&No\\
 \textcolor{red}{\textbf{AS2}}&Yes&No&Yes&No&Yes&No&No \\
 
 \textcolor{red}{\textbf{AvL}}&Yes&Yes&Yes&No&No&No&No\\
 
 \textcolor{red}{\textbf{CLE}}&Yes&Yes&No&No&No&NFW&NFW\\
 
 \textcolor{red}{\textbf{CLN}}&Yes&Yes&No&No&No&NFW&NFW\\
 
 \textcolor{red}{\textbf{SG1}}&Yes&Yes&Yes&No&No&No&No\\
 
 \textcolor{red}{\textbf{SG2}}&Yes&Yes&Yes&No&No&No&No\\
 
 \textcolor{red}{\textbf{PCS}}&Yes&Yes&No&No&No&No&No \\
 
 \textcolor{red}{\textbf{PFS}}&Yes&Yes&No&No&No&No&No \\
 \bottomrule
 \end{tabular} 
  \label{table:appendix_table_2}
\end{table*}
\clearpage
\onecolumn
\begin{flushleft}
\section{Appendix B: recovered mass distributions and residuals}
\end{flushleft}
\setcounter{figure}{0} \renewcommand{\thefigure}{B\arabic{figure}}
\begin{figure*}
 \centering
\includegraphics[trim = -5mm 16mm 0mm 25mm, clip, width=0.95\textwidth]{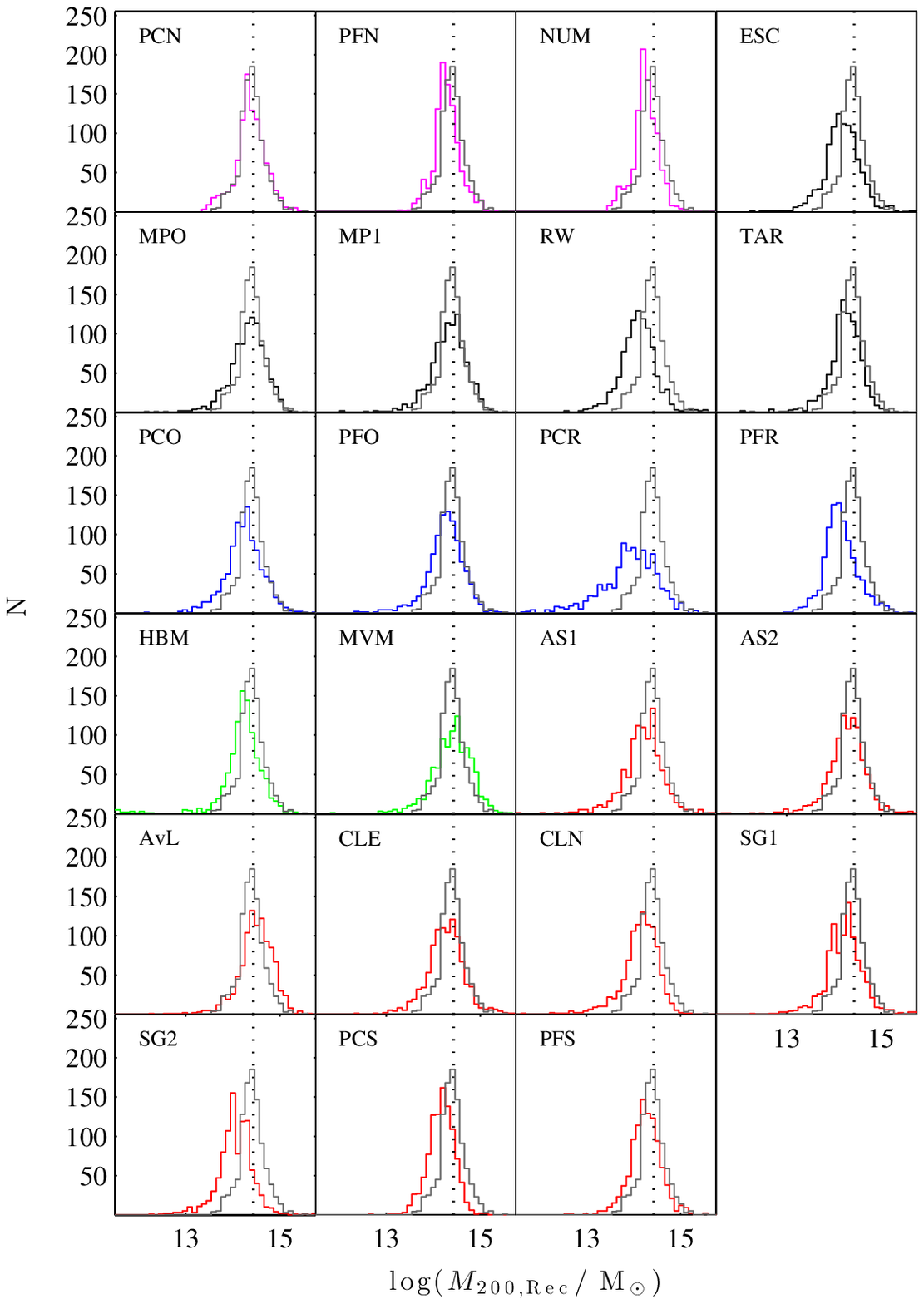}
 \caption{Recovered mass distributions when the group/cluster membership
 is not known. The black dotted line represents the mean of the true
 mass distribution and the grey distributions on each subplot
 represent the true mass distributions.} 
\label{fig:phase_1_mass_hists}
\end{figure*}
\begin{figure*}
 \centering
\includegraphics[trim = -2mm 16mm 0mm 25mm, clip, width=1.0\textwidth]{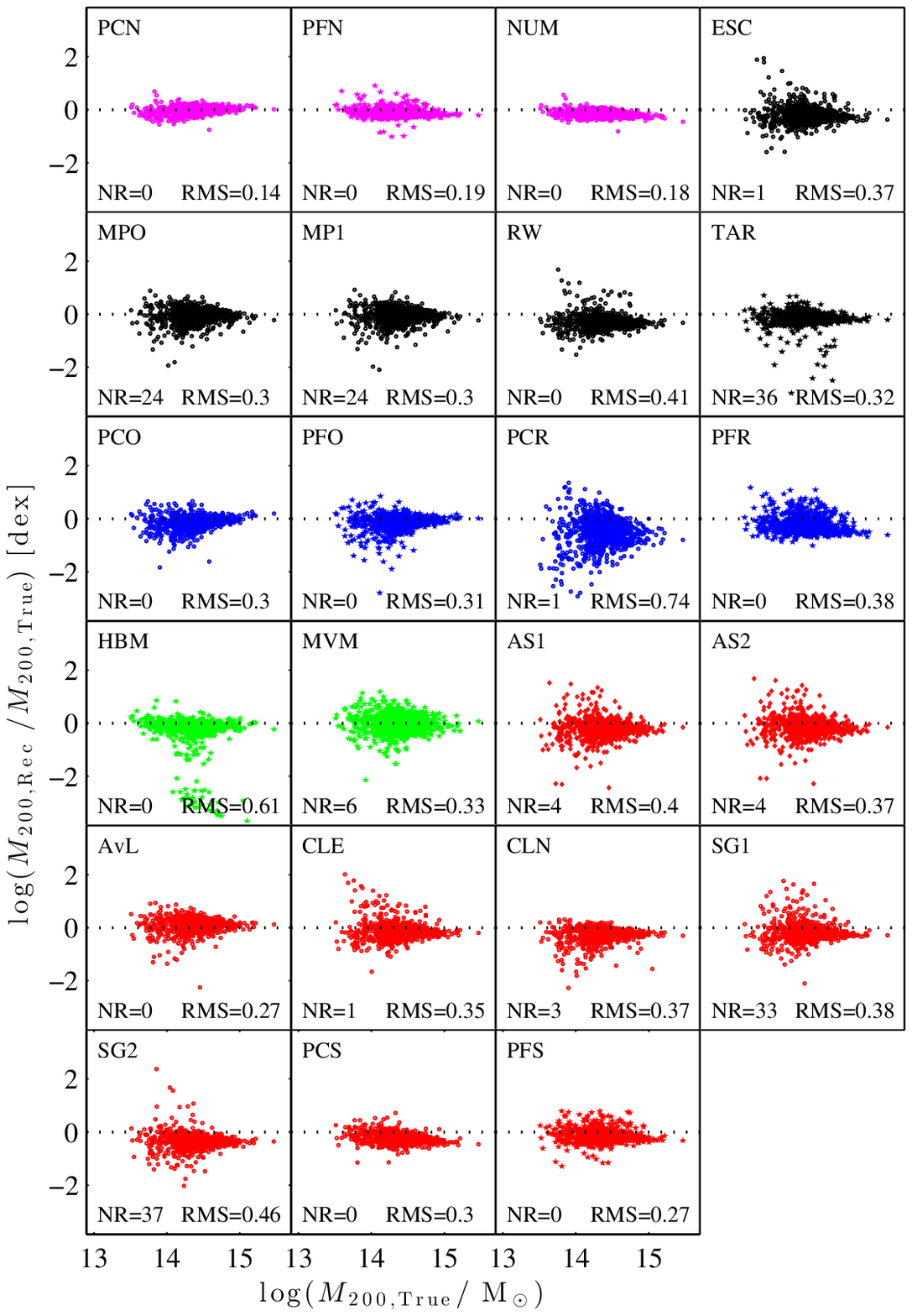}
 \caption{Mass bias versus true mass when
 the group/cluster membership is not known. The black dotted line represents a residual of zero. `NR' in the legend represents groups/clusters
 that are not recovered because they are found to have very low
 ($\rm < 10^{10} \,M_{\odot}$) or zero mass.} 
\label{fig:phase_1_mass_residual}
\end{figure*}
\begin{figure*}
 \centering
 \includegraphics[trim = -5mm 16mm 0mm 25mm, clip, width=1.0\textwidth]{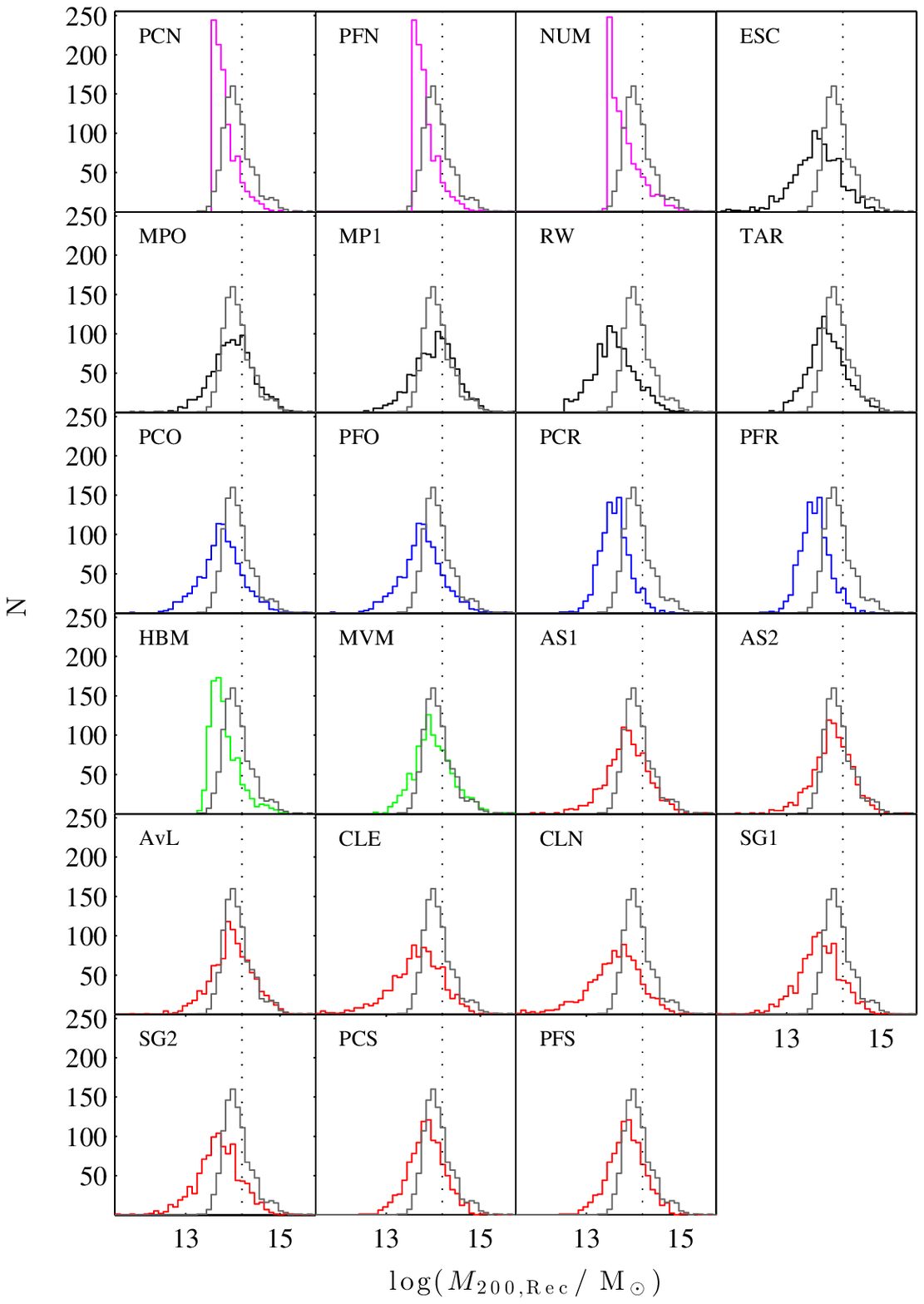}
 \caption{Recovered mass distributions when the group/cluster
 membership is known. The black dotted line represents the mean of
 the true mass distribution and the grey distributions on each
 subplot represent the true mass distributions.} 
\label{fig:phase_1_mass_hists_truemembership}
\end{figure*}
\begin{figure*}
 \centering
 \includegraphics[trim = -5mm 16mm 0mm 25mm, clip, width=1.0\textwidth]{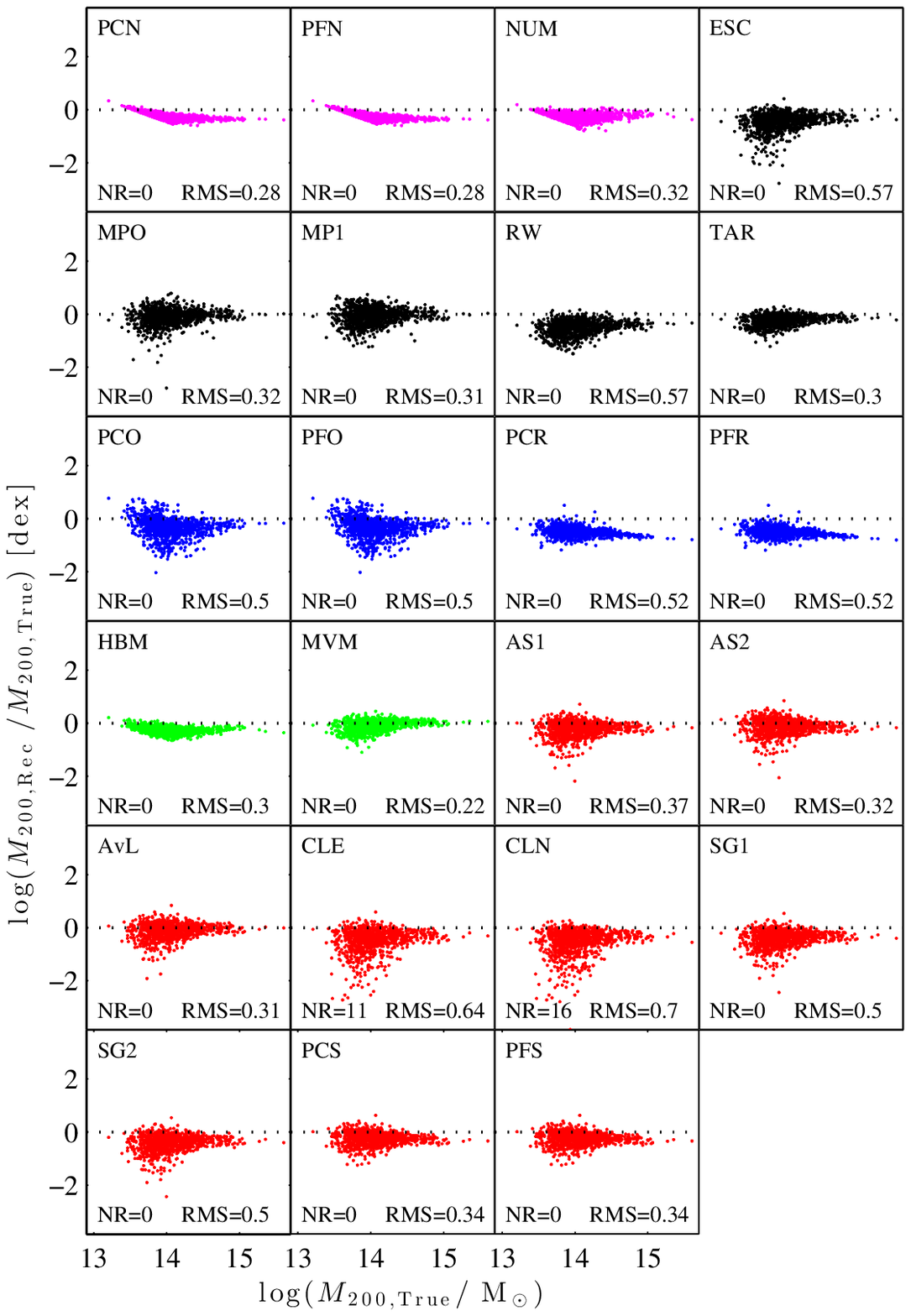}
 \caption{Mass bias versus true mass when
 the group/cluster membership is known. The black dotted line represents a residual of zero. `NR' in the legend represents groups/clusters
 that are not recovered because they are found to have very low
 ($\rm < 10^{10} \,M_{\odot}$) or zero mass.} 
\label{fig:phase_1_mass_known_residual}
\end{figure*}
\clearpage
\onecolumn
\begin{flushleft}
\section{Appendix C: recovered number of galaxies distributions and residuals}
\end{flushleft}
\setcounter{figure}{0} \renewcommand{\thefigure}{C\arabic{figure}}
\begin{figure*}
 \centering
 \includegraphics[trim = -5mm 15mm 0mm 25mm, clip, width=0.95\textwidth]{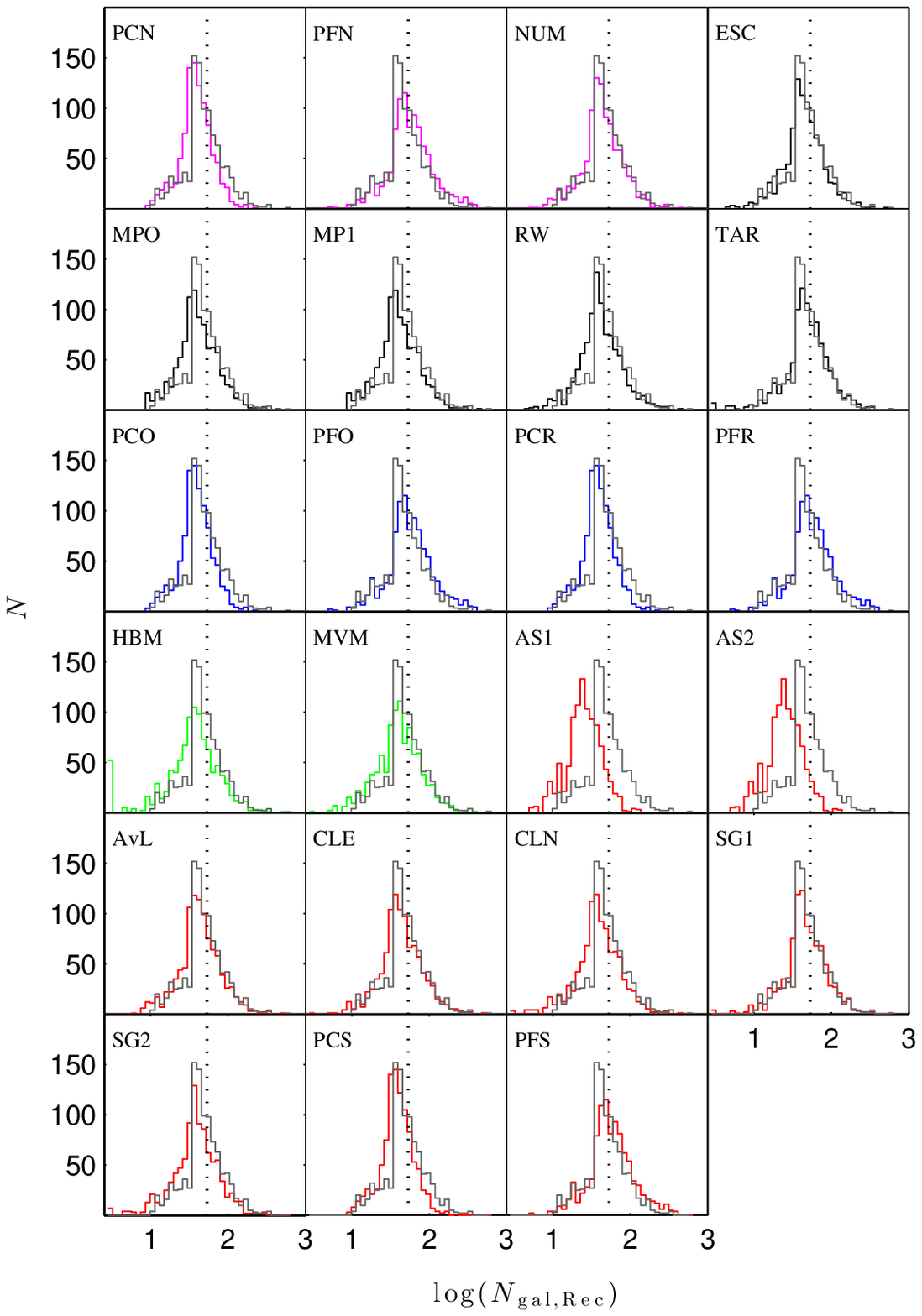}
 \caption{Distributions of the recovered number of galaxies associated
 with each group/cluster when the membership is not known. The grey distribution on each subplot
 represents the true richness distribution of the
 groups/clusters. The black dotted line presents the mean of this
 input richness distribution.} 
\label{fig:phase_1_ngal_hists}
\end{figure*}
\begin{figure*}
 \centering
 \includegraphics[trim = -5mm 14mm 0mm 25mm, clip, width=1.0\textwidth]{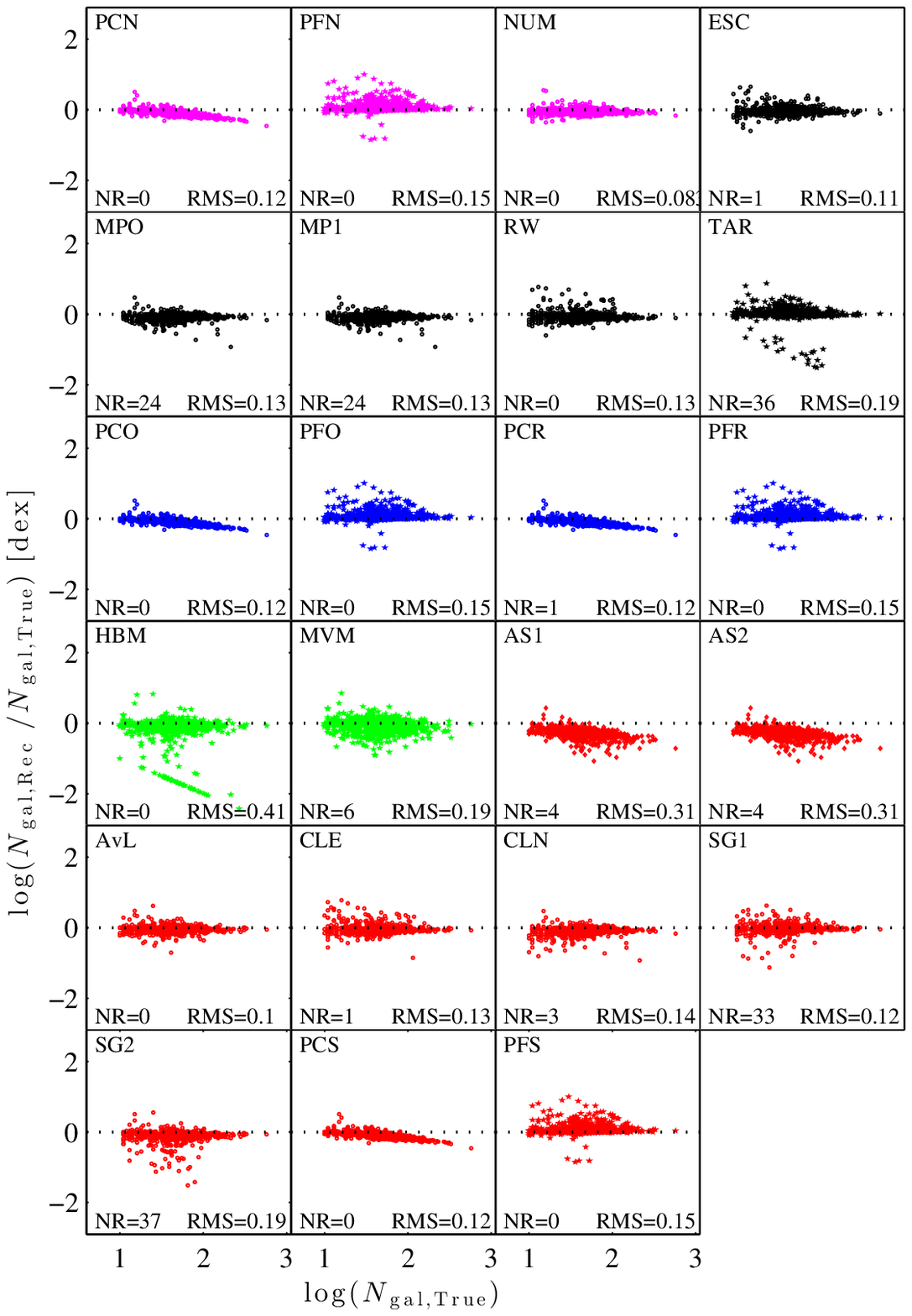}
 \caption{Richness bias versus true richness when the group/cluster membership is not known. The black dotted line represents a residual of zero. `NR' in the legend represents groups/clusters that are not recovered because
 they are found to have very low ($\rm < 10^{10} \rm M_{\odot}$) or zero mass.} 
\label{fig:phase_1_ngaltrue_nalcat_residual}
\end{figure*}
\label{lastpage}
\end{document}